\documentclass[pre,amsmath]{revtex4}
\usepackage[pdftex]{graphicx} 
\usepackage{color}
\usepackage{bm}
\usepackage{epsfig}
\usepackage{latexsym}
\usepackage{nameref,hyperref}
\usepackage{epstopdf}
\usepackage{float}
\begin{document}
\newcommand{\be}{\begin{equation}}
\newcommand{\ee}{\end{equation}}
\newcommand{\bea}{\begin{eqnarray}}
\newcommand{\eea}{\end{eqnarray}}
\newcommand{\RN}[1]{\textup{\uppercase\expandafter{\romannumeral#1}}}
	
\title{Persistent exclusion process with time-periodic drive}
\author{Deepsikha Das and Sakuntala Chatterjee}
\affiliation{Department of Physics of Complex Systems, S. N. Bose National Centre for Basic Sciences, Block JD, Sector 3, Salt Lake, Kolkata 700106, India.}
\begin{abstract}
We study a persistent exclusion process with time-periodic external potential on a one-dimensional periodic lattice through numerical simulations. A set of run-and-tumble particles move on a lattice of length $L$ and tumbling probability $\gamma \ll 1$ and interact among each other via hardcore exclusion. The effect of the external potential has been modeled as a special site where the tumbling probability is $1$. We call it a "defect" site and move its location along the ring lattice with speed $u$. In the case of $\gamma=0$ the system goes to a jammed state when there is no defect. But introduction of the moving defect creates a strongly phase-separated state where almost all active particles are present in a single large cluster, for small and moderate $u$. This striking effect is caused by the long-range velocity correlation of the active particles, induced by the moving defect. For large $u$, a single large cluster is no longer stable and breaks into multiple smaller clusters. For nonzero $\gamma$ a competition develops between the timescales associated with tumbling and defect motion. While the moving defect attempts to create long-range velocity order, bulk tumbling tends to randomize the velocity alignment. If $\gamma$ is comparable to $u/L$, then a relatively small number of tumbles take place during the time the moving defect travels through the entire system. In this case, the defect has enough time to restore the order in the system and our simulations show that the long-range order in velocity and density survive for $\gamma$ values in this range. As $\gamma$ increases further, long range order is destroyed and the system develops multiple regions of high and low density. We characterize the density inhomogeneity in this case by measuring subsystem density fluctuations and present a heatmap in the $\gamma$-$u$ plane showing the regions with most pronounced density inhomogeneities.
\end{abstract}
\maketitle
	
\section{Introduction} 
\label{intro}

Active matter systems have received a lot of recent research attention \cite{Ramaswamy_2017, schweitzer2003brownian, bechinger2016active}. Such systems describe a wide variety of natural phenomena, such as swimming of a school of fish, flight of a flock of birds, and migration of a bacterial colony \cite{schweitzer2003brownian, gregoire2004onset, toner1995long, vicsek2012collective, toner1998flocks, tu1998sound}. The motions observed in all these systems have one key characteristic: each individual member of the population consumes energy to perform directed motion. Thus, active matter systems are intrinsically  nonequilibrium even at the level of a single particle. To understand the behavior of such systems a number of theoretical models are used. Examples include Vicsek's model, active Brownian particles, run-and-tumble particles, etc \cite{vicsek1995novel, ginelli2016physics, fily2012athermal, redner2013structure, tailleur2008statistical, slowman2016jamming, cates2013active}. In these models, while the "activity" of a single particle is modeled as self-propulsion, the interaction among the particles can be of different types. In some cases the particles show an aligning interaction, where the direction of their motion tends to get aligned with that of their neighbors \cite{vicsek1995novel, vicsek2012collective}. In some other cases, the interaction affects the speed of the particles, making them move slower in crowded regions. These models can successfully capture many collective phenomena observed in active matter systems, such as coarsening and aggregation \cite{sepulveda2016coarsening, levis2014clustering, stenhammar2014phase, mognetti2013living, redner2013structure, suma2014motility}, pattern formation on mesoscopic scales \cite{cates2010arrested, geyer2019freezing, solon2015pattern}, and giant density fluctuations \cite{marchetti2013hydrodynamics, vicsek2012collective, bialke2015active}.

One particularly simple form of interaction among particles is excluded volume interaction which does not allow two particles to overlap. Nonaligning active particles with excluded volume can lead to aggregation because of a competition between self-propulsion of the particles and their mutual exclusion. Active motion of hard-core particles on a one-dimensional lattice was studied for the first time in \cite{soto2014run}. This model, also known as persistent exclusion process (PEP) consists of run-and-tumble particles moving on a one-dimensional lattice with the constraint that no two particles can occupy the same lattice site. Depending on the tumbling rate, i.e., the rate at which the particles switch their run direction, and the particle density, this system shows clustering among the particles which gives rise to density inhomogeneity in the system. In subsequent generalization of the PEP, short-range attractive interaction among the particles was included and it was shown that, even without any alignment interaction, motile clusters can form in the system \cite{barberis2019phase, gutierrez2021collective}. In \cite{dandekar2020hard} using a mapping between mass transfer models and PEP, hydrodynamic coefficients of diffusivity and conductivity were calculated analytically.

In this paper, we ask what happens when a set of interacting run-and-tumble particles is subjected to external potential. This question has been studied in great detail in the context of a single run-and-tumble particle, which either changes its run speed or its tumbling rate in response to an external field. The most common example can be found in bacterial chemotaxis \cite{berg2004coli}, where certain bacteria such as {\sl Escherichia coli, Salmonella typhimurium, and Basillus subtilis,} modulate their tumbling rate in the presence of a chemical (nutrient) gradient in their environment in order to move up the gradient \cite{dev2018optimal, mandal2021effect}. Using the same mechanism, a bacterial colony also shows migration toward nutrient-rich regions \cite{eisenbach2004chemotaxis}. Here, we consider the situation where the external potential is time periodic in nature, and explore the behavior of hard-core run-and-tumble particles in that potential. To be more specific, we consider a PEP in one dimension where hard-core particles, moving either rightward or leftward on a periodic lattice,  switch their directions with a small probability $\gamma$. The external potential is modeled as a "defect site" where the local tumbling probability is different from $\gamma$ and takes the value $1$. This means that tumbling events occur with small probability everywhere else on the lattice, but if a particle is present on the defect site, it tumbles with all probability. Due to the time-periodic nature of the potential, the defect site is not static but moves through the lattice with uniform velocity $u > 0$. We are mainly interested in how introduction of the moving defect affects the long-time behavior of the active system. In a number of earlier studies the effect of such a periodically moving defect was investigated in detail \cite{chatterjee2014interacting, chatterjee2016symmetric, das2023optimum} for a system of interacting particles but no activity was considered there. In this paper, we find out what happens when activity is included. To the best of our knowledge a periodically driven active matter system has not been studied before.

Our Monte Carlo simulations show that the moving defect strongly affects the steady state of the system. The effect is most dramatic for the limit when $\gamma$ vanishes. A PEP with $\gamma=0$ means the hardcore particles are not able to switch their directions. Two oppositely moving particles moving toward each other occupy neighboring sites, and no more movement is possible. In this case it is easy to see that the system gets arrested in a jammed state where particles form small immobile clusters. The introduction of the moving defect makes it possible for the system to exit the jammed state since it offers the particles a chance to switch their velocity. For small or moderate defect velocity, we find that the system goes to a strongly ordered state, where almost all active particles are present in a single large cluster. We show that this clustering is caused by the defect-induced long-range order in the velocity of the active particles. For large defect velocity, a single large cluster cannot survive anymore and breaks up into multiple clusters of smaller size, which are separated by empty stretches of the lattice. We quantitatively characterize the order present in the system by measuring two point velocity correlations and two point density correlations and examine their scaling with system size $L$ for small and large $u$ values.

For nonzero $\gamma$ we find a competition between tumbling timescale and defect movement timescale. The random tumbling events at the bulk tend to destroy the long-range velocity order created by the moving defect. The result of this competition depends on which process is faster. For a slowly moving defect, if $\gamma$ is small enough such that not too many tumbles take place during the time $L/u$ needed by the defect to travel through the entire ring lattice and come back to the same point, then the loss of order due to tumbles happens slowly and the defect has enough time to restore the order. In this case, we do find long-range velocity and density order in the system. The range of $\gamma$ values for which this is observed depends on $L$, as follows from the above argument. As $\gamma$ increases further, more and more tumbles take place and eventually it becomes impossible for the defect to restore the order. The system cannot support a single large particle cluster anymore and in steady state multiple regions of high and low density develop. To characterize the density inhomogeneity in this case we measure subsystem density fluctuations and study its $u$ dependence. We present a heat map in the $\gamma$-$u$ plane showing the regions with most pronounced density inhomogeneities. We also measure the density and velocity profile created by the moving defect and find strong, qualitative changes with $u$. We explain how this behavior showcases the competition between bulk tumbling and defect movement.

In the next section we present our model. In Sec. \ref{g0} we present our simulation results for $\gamma =0$ and show how a clear ordered state is obtained for small and moderate $u$ values. In Sec. \ref{gn0} we consider finite $\gamma$ and present different measures of the short-range density inhomogeneities present in the system.  In Sec. \ref{orderdisorder} we consider $\gamma$ values around $u/L$  where long-range order like $\gamma=0$ is restored, and explicitly show how such order slowly disappears with tuning $\gamma$. Our conclusions are presented in Sec. \ref{sec:con}.

\section{Model description} 
\label{model} 
We consider an active lattice gas showing a persistent exclusion process. The model consists of $N$ hard-core particles with run-and-tumble motion on a ring of $L$ sites. Each site can have a maximum of one particle and each particle has a velocity associated with it, which can be either $+1$ or $-1$. We often refer them as "right-mover" or "left-mover", respectively. A particle hops to its neighboring site, in the direction of its velocity, if that neighboring site is empty. With a small probability $\gamma \ll 1$ a particle can tumble, when its velocity switches sign. The external drive causes an increase in the tumbling frequency. In our model we incorporate it as a special site, termed a "defect" site, where the tumbling probability is $1$, much larger than the bulk value $\gamma$. This defect site moves through the periodic lattice with uniform velocity $u$, such that it comes back to each position after a time period $L/u$, which makes the drive time periodic. Without any loss of generality, we consider $u$ to be positive.

\begin{figure}[h!]
\includegraphics[scale=0.4]{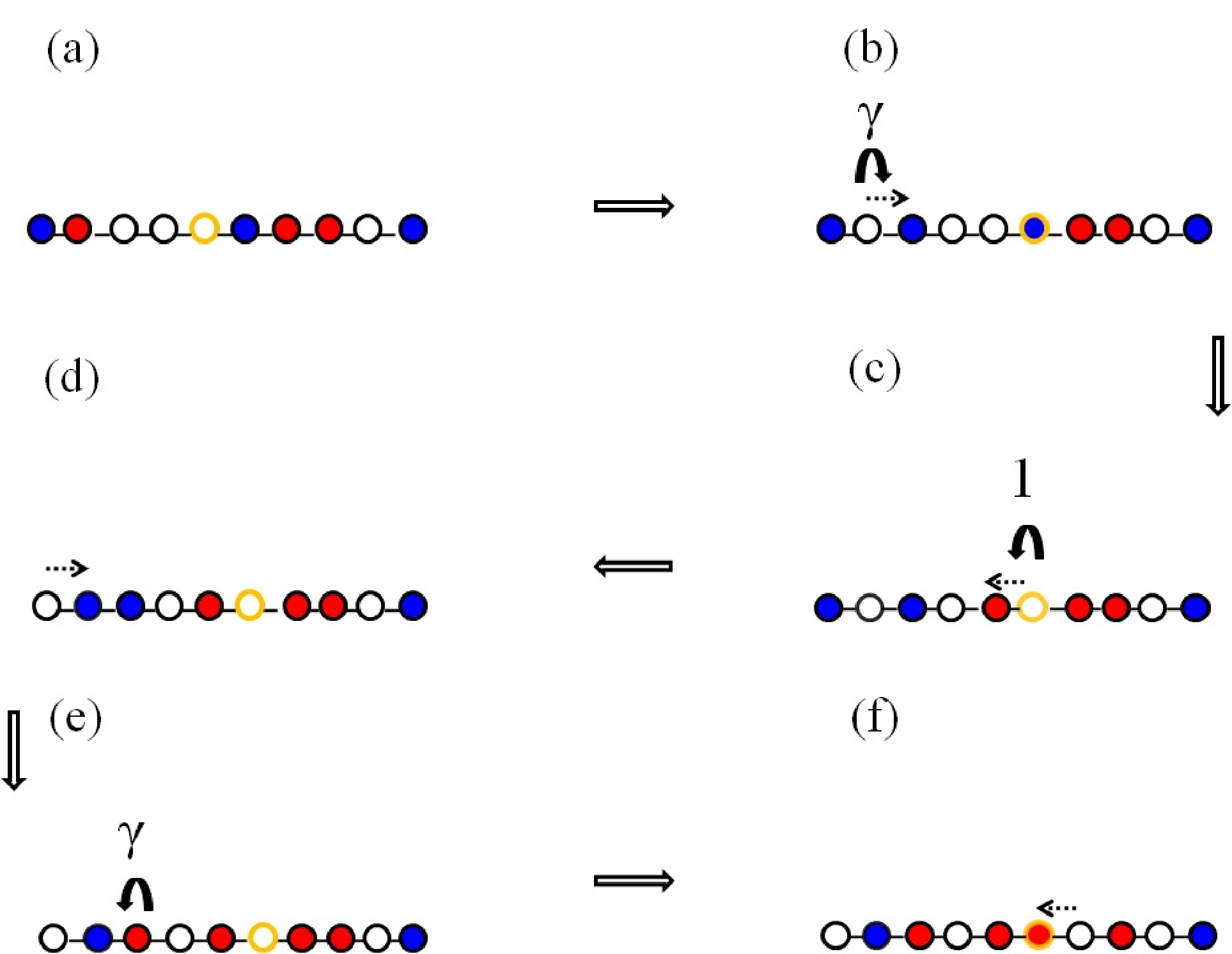}
\caption{ Schematic description of transitions in a representative configuration. The defect site is shown as an orange circle and the bulk sites as black circles. If a site is occupied by a right-mover (left-mover), the circle is filled with blue (red), and is left empty otherwise. (a) Initial configuration shows defect sitting on an empty site. (b) One left-mover at the bulk tumbles, becomes a right-mover and moves to the empty site on its right. Meanwhile, the defect site also changes its position. (c) The right-mover at the defect site tumbles, becomes a left-mover and hops leftward. (d) A right-mover at the bulk hops to its next site. (e) A right-mover at the bulk tumbles to become a left-mover but cannot move since the left neighbor is already occupied. (f) A left-mover moves into the defect site. } \label{themodel}
\end{figure}

We perform Monte Carlo (MC) simulations on this model. We use random sequential updates and each MC time step consists of $L$ such update trials. During an update we choose a lattice site at random and if it is occupied by a particle, then we switch its velocity sign with probability $\gamma$ or $1$, depending on whether the chosen site is a bulk site or a defect site. After the velocity switching attempt, the particle attempts to move by one site, in the direction of its current velocity. The move is successful only if the destination site is empty. After residing for a time $\tau=1/u$ at a particular site, the defect moves to the next site. In Fig. \ref{themodel} we have pictorially shown the transitions. 

Unless mentioned otherwise, we perform all measurements stroboscopically, when the defect has finished its residence time $\tau$ at one site and is about to move to the next one. All our measurements are performed in the long-time limit, when the system has reached a time-periodic steady state. Our data are averaged over a steady-state ensemble, and we have used a minimum ensemble size of $10^6$ for all our data.

\section{No tumbling in the bulk: \boldmath $\gamma=0$} \label{g0}

\begin{figure}[h!]
\includegraphics[scale=0.4]{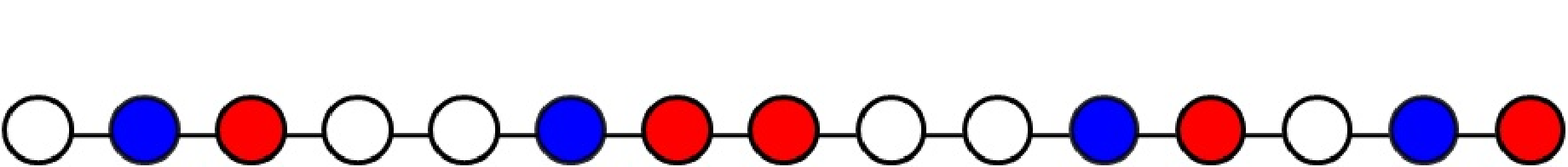}
\caption{An example of a jammed configuration reached at $\gamma=0$ in absence of any defect. The left-movers are shown in red, right-movers in blue.}
\label{fig:jam}
\end{figure}

For $\gamma =0$ no tumbling takes place in the system if no external drive is present. In this case, starting from an initial configuration, the particles move rightward or leftward, depending on their velocities but movement stops whenever a particle finds its destination site occupied by another particle with opposite velocity. Thus, a right-left pair gets stuck and all right-movers (left-movers) appearing on the left (right) of this pair pile up to form a cluster. This leads to a "jammed" state, where each particle is immobile and part of a cluster of minimum size $2$, with no isolated particle left in the system. In Fig. \ref{fig:jam} we show such a configuration. The final jammed configuration in which the system gets stuck- depends on the initial arrangement of right- and left-movers. No long-range order is present in the system in this case. 

However, when the time-periodic external drive is present, it induces tumbling and the particles do not remain stuck anymore. The resulting steady state is very different from a jammed state. We find that the presence of the time-dependent external potential or moving defect gives rise to long-range order in the velocity-velocity correlation of the active particles, owing to which the particles form a macroscopic cluster in the system. In steady state, almost all the particles are found to be present in a single large cluster. This clustered phase coexists with a sparsely populated region which is either empty or contains very few particles. The phase separation of particles and holes is most pronounced for small $u$ when the defect moves slowly through the system, giving rise to a single particle cluster made of all the particles, coexisting with a single hole cluster. But as $u$ increases,  the particles tumble much more frequently and the long-range order in velocity-velocity correlation is destroyed. A single large cluster cannot form in this case. Instead, the system contains multiple clusters and the phase separation is less complete than in the small-$u$ case. In the remaining part of this section we discuss these effects in detail.

\subsection{Time evolution toward ordered state}
\label{sec:g0t}
	
\begin{figure}[h!]
\includegraphics[scale=1.4]{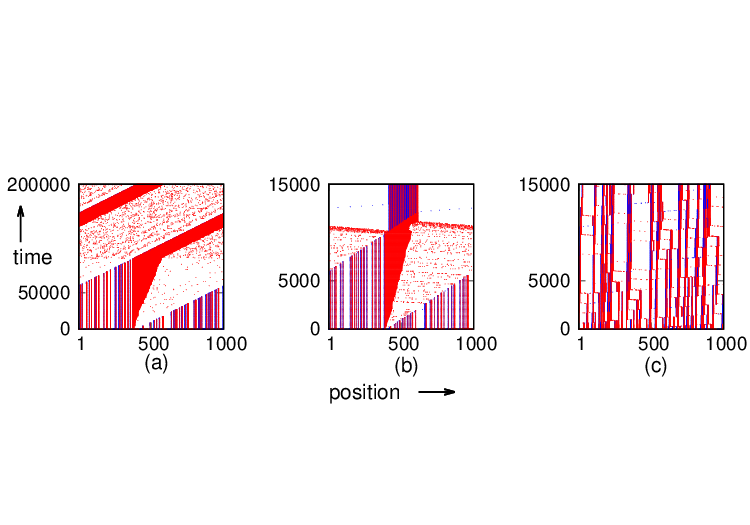}
\caption{Time evolution toward phase-separated state. Right-movers are shown in blue and left-movers are in red. (a) $u=0.01, (b) 0.1, and (c) 1$. $\rho=0.2$, $L=1000$ here.}
\label{fig:g0t}
\end{figure}

	Starting from a randomly disordered configuration, under the action of the time-periodic drive, the system evolves toward a state where the particles form macroscopic clusters. Like any other exclusion model \cite{nagar2008boundary, chatterjee2023counterflow,chacko2024clustering} we define a particle cluster as a chain of successive occupied sites, with two empty sites at the two ends. Note that for a cluster to be stable, we must have a right-mover (left-mover) at its left (right) edge. In Fig. \ref{fig:g0t} we show the time evolution of an initial random configuration for different values of $u$. In all the cases, the active particles form large clusters, but for small $u$ we find quite often a single large cluster is present containing all the particles in it. We find that such large cluster formation is a consequence of the long-range velocity ordering induced by the defect movement. We explain the mechanism below.  

Since the defect is moving from left to right, if it encounters a right-mover, it switches its velocity from $+1$ to $-1$, and soon this newly transformed left-mover moves behind the defect. On the other hand, if a left-mover switches its velocity from $-1$ to $+1$ under the action of the defect, then it moves in front of the defect. This gives the defect an opportunity to catch up with the particle again and switch its velocity once more to $-1$ and now this particle can move behind the defect as before. This shows that as the defect moves through the system, it leaves a trail of left-movers behind it. These left-movers keep moving leftward until the leftmost particle among them encounters a right-mover and it gets stuck, following which all the other left-movers pile up behind it and thus a large particle cluster is formed. For small $u$, the right-movers can exist only in front of the defect, but as $u$ becomes large, the residence time $\tau =1/u$ of the defect at any given location is small, and it is possible that, after switching the velocity of a particle from $-1$ to $+1$, the defect moves to the next site, before the newly formed right-mover manages to move. Therefore, for large $u$ the defect leaves a trail of both left-movers and right-movers behind it, and the velocity alignment is not as perfect as the one found for small $u$. As a result, for large $u$, the particles form multiple clusters, each containing a right-mover at its left edge, and a left-mover at its right edge.

In the remaining part of this section we characterize the steady-state properties of the ordered state observed for small and large $u$.

\subsection{Clustering in steady state} 
\label{csdg0}

\begin{figure}[H]
	\centering
	\includegraphics[scale=0.6]{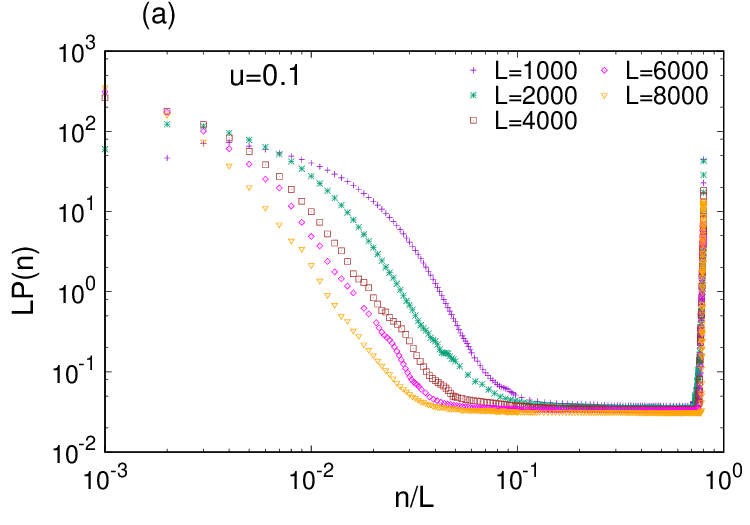}
	\includegraphics[scale=0.6]{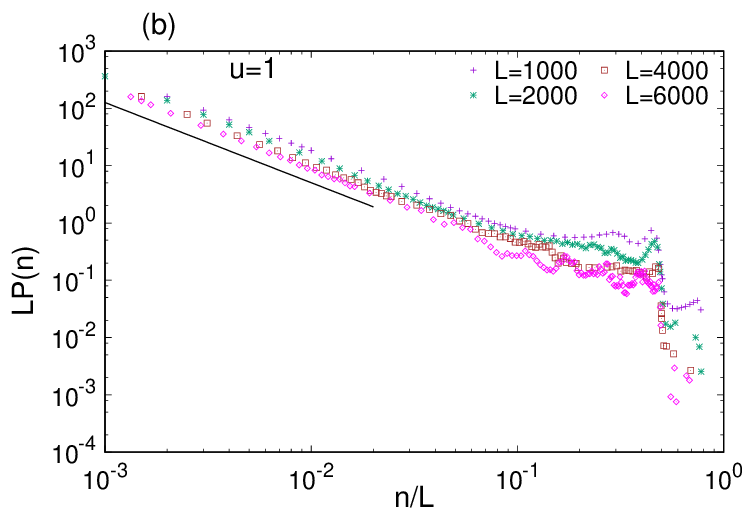}
\caption{Probability distribution $P(n)$ for hole clusters scaled by $L$, plotted against $n/L$ for $u=0.1, 1$ and $\rho=0.2$. For small $u$, $P(n)$ shows a  distinct large peak appearing at $n=L-N$. When $u$ is large, $P(n)$ decays as a power law with an exponent $\simeq 1.4$ (solid line). These data are averaged over $10^7$ configurations.}
\label{csd0} 
\end{figure}

As shown in Fig. \ref{fig:g0t}, starting from a homogeneous initial state, the system forms large particle clusters in the long-time limit. Note that for these hard-core particles, the formation of large particle clusters automatically implies the formation of large hole clusters. We measure the probability distribution of hole cluster size in steady state. A hole cluster of size $n$ is defined as an uninterrupted sequence of $n$ empty sites, bounded by two occupied sites at the two ends. Let $P(n)$ be the probability to observe a hole cluster of size $n$ in a steady-state configuration. In Figs. \ref{csd0}(a) and \ref{csd0}(b) we present the data for $P(n)$ vs $n$ for a small and a large $u$, for different values of system size $L$. After scaling the data with $L$, we find that when $u$ is small, the scaled distribution shows a large peak for $n/L = (1-\rho)$, with $\rho$ being the average particle density in the system. This means that there is a very large probability to find all the holes present in a single cluster for small $u$. However, when $u$ is large, the probability of a single large hole cluster decreases significantly. $P(n)$ in this case shows a power-law decay with $n$, with an exponent $\simeq 1.4$. As explained in Sec. \ref{sec:g0t}, for fast movement of the defect, the particles do not have a perfect velocity ordering, which is necessary to form a single large cluster. In steady state the system contains multiple particle clusters, as shown in Fig. \ref{fig:g0t}. Therefore, finding a single hole cluster also becomes less likely.  

\begin{figure}[H]
\centering
\includegraphics[scale=0.7]{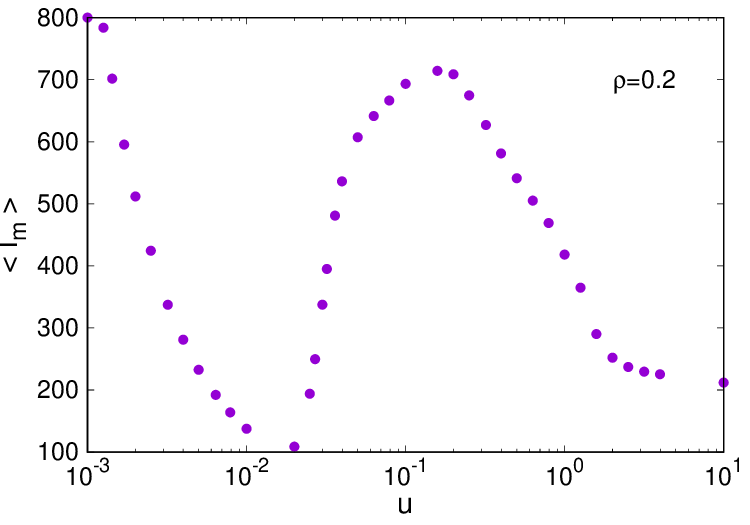}
\caption{ Average size of the largest hole cluster, $\langle l_m \rangle$, plotted against $u$ for $\rho=0.2$ and $L=1000$. The position of the peak matches reasonably well with our analytical estimation $N/(L-N) = 0.25$.  }    
\label{lmaxu}
\end{figure}

To quantitatively study how the clustering property changes with defect velocity $u$, we plot the average size of the largest hole cluster, $\langle l_m \rangle$, as a function of $u$ in Fig. \ref{lmaxu}. Our data show interesting nonmonotonic variation, where $\langle l_m \rangle$ first decreases with $u$, reaches a minimum, then increases again to reach a peak and finally decreases with $u$ when $u$ becomes large. To explain this behavior we consider how the largest hole cluster gets affected when the moving defect passes through the largest particle cluster. 

For very small $u$, all the particles are present in a single large cluster, and there is a single hole cluster containing all empty sites. As explained in Sec. \ref{sec:g0t}, under the action of the moving defect, one after another particle in the cluster jumps behind the defect as left-movers. These left-movers get detached from the left edge of the cluster, move through the empty stretch of the lattice and rejoin the particle cluster at the right edge. The residence time of the defect $\tau = 1/u$ is large enough such that before the defect moves to its next position, the particle(s) which escaped had sufficient time to cross the empty region and rejoin the cluster. In our stroboscopic measurements, performed just before the defect moves to a new position, we almost always observe the defect at the left edge of the single particle cluster containing all particles and  $\langle l_m \rangle \simeq (1-\rho)L$. As the defect moves, the left edge shrinks and the right edge extends, as described above. As $u$ increases, $\tau$ becomes small such that it may not be always possible for the detached particles to travel through the entire empty stretch of the lattice before the defect moves. In this case, during our measurements one or more left-movers may be present in that otherwise empty region. This breaks the hole cluster and $\langle l_m \rangle$ falls below  $(1-\rho)L$. As $u$ increases further, the probability to find particle(s) in the middle of the empty stretch is higher, and $\langle l_m \rangle$ decreases. Even for this range of $u$, it is most likely to find the defect at the left edge of the particle cluster.

\begin{figure}[H]
\centering
\includegraphics[scale=0.4]{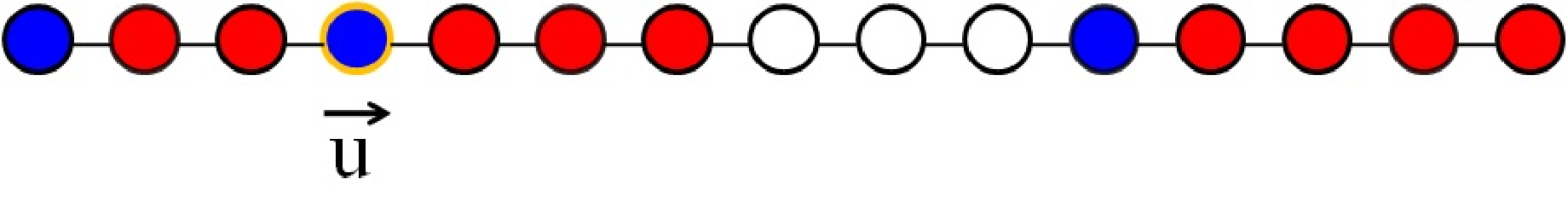}
\caption{ A configuration showing the largest particle cluster fragmented into two smaller clusters separated by a few holes.  Such a scenario arises for a moderate $u$, because of a reduced residence time of the defect. Here the defect leaves a right-mover
behind it and moves on through the cluster. Blue (red) denotes right-mover (left-mover).}
\label{fig:fragment}
\end{figure}

However, as $u$ keeps increasing, the downward trend of $\langle l_m \rangle$ is interrupted by another effect. As explained in Sec. \ref{sec:g0t}, for larger $u$ values, a right-mover can also find itself behind the defect. Such a right-mover now becomes the stable left edge of the cluster, and no other particle can escape through the left edge anymore.  Meanwhile, after leaving the left edge, the defect moves through the bulk of the cluster, and although under the action of the defect the particles in the cluster switch the sign of their velocities, this sign reversal produces no effect since these particles cannot leave the cluster. Now that no particles are able to leave the cluster, the empty stretch of the lattice is not fragmented anymore by the presence of particles and $\langle l_m \rangle$ starts increasing again. Note that both right-movers and left-movers are now present inside the particle cluster. This makes it possible for a small number of holes to survive in this region for a significant amount of time (in the presence of only left-movers the holes would have zipped through the cluster). In Fig. \ref{fig:fragment} we show a representative local configuration. This effect is also visible when we measure the probability distribution of particle clusters (see Fig. \ref{csdpg0u0p1}), which shows an intermediate peak for moderate cluster size when $u$ is in this range. Finally, for very large $u$, large particle clusters cannot form anymore due to poor velocity ordering by a fast-moving defect. Moreover the defect switches velocity of the particles so frequently- that the particle clusters are no longer stable and evolve continuously. This in turn does not allow a large hole cluster to survive resulting in multiple small hole clusters which are present throughout the system. This causes  $\langle l_m \rangle$ to decrease with $u$ once more after it reaches a peak. 

It may be useful to estimate an order of magnitude of $u$ when a single large particle cluster becomes unstable and multiple smaller clusters begin to form throughout the length of the system. Note that this state is distinctly different from the one when a small number of holes may enter the large particle cluster and break the cluster into a few smaller segments (Fig. \ref{fig:fragment}). In such a state, all the particles still occupy a specific portion of the lattice and there is a large hole cluster containing almost all the empty sites in the system. On the contrary, the state obtained at large $u$ has multiple particle clusters present throughout the lattice [as shown in Fig. \ref{fig:g0t}(c)]. To estimate $u$ which gives rise to such states, first consider a single large particle cluster containing all $N$ particles. The time required for the defect to move through this cluster from the left edge to the right edge is $\tau_d \sim N/u$. The time needed for the left-mover which got detached from the cluster's left edge- to travel through the empty region of length $(L-N)$ and rejoin the cluster's right edge is $\tau_L \sim (L-N)/v $, where particle velocity $v$ has magnitude unity in our model. If $\tau_d < \tau_L$ then before the detached left-movers get a chance to join the cluster, they are intercepted by the moving defect which now switch their velocity and may form smaller local clusters consisting of these detached particles. Putting $\tau_d \sim  \tau_L$ gives us $u^\ast \sim N/(L-N)$, which estimates the defect velocity $u^\ast$ beyond which the system cannot support a single large cluster anymore, and $\langle l_m \rangle$ starts falling off its peak. In Fig. \ref{lmaxu} we have used $\rho = 0.2$ for which $u^\ast = 0.25$ which shows good agreement with our numerical data. As $\rho$ increases, $u^\ast$ also increases. Note that our simple estimate of $u^\ast$ relies on the picture that for $u \lesssim u^\ast$ a single large cluster is present in the system. But this breaks down for large $u$ due to loss of velocity ordering which makes the clusters unstable anyway. In Fig. \ref{fig:lmax_u_rho} in Appendix \ref{app:lm} we compare our estimate of $u^\ast$ with numerical data and find deviation for large $\rho$ which predicts large $u^\ast$.

\begin{figure}[h!]
\centering
\includegraphics[scale=1.1]{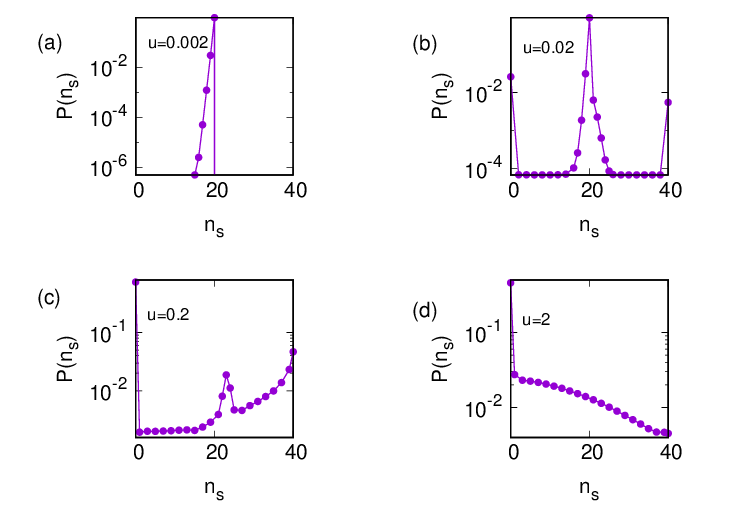}
\caption{Probability distribution of particle numbers $P(n_s)$ in a box of size $n_s=40$ centered at the defect site. Here $\rho=0.2, L=1000$. The distribution changes its behavior with increase in $u$, which is consistent with the mechanisms of defect-induced evolution of the largest particle cluster over different $u$ regime, as explained in the above text.}
\label{Pls}
\end{figure}

To further test our arguments, we numerically verify the position of the defect relative to the particle cluster for different ranges of $u$ values. As follows from the above argument, for small $u$ we expect the defect to be localized at the left edge of the cluster and for moderately large $u$ it moves to the bulk, whereas for very large $u$ it crosses the whole cluster and can even be found in empty regions of the lattice. To test this picture, we define a box of size $s$ around the defect and count the number of particles in that box. Let $P(n_s)$ be the probability to find $n_s$ particles in the box, with $0 \leq n_s \leq s$. If the defect is at the left edge of the cluster, then one expects $n_s =s/2$, whereas for the defect in the bulk of the cluster $n_s =s$ is expected. In Fig. \ref{Pls}, we plot $P(n_s)$ for different $u$ values and for a fixed $s=40$, while we consider $\rho=0.2$. Indeed, for small $u$, $P(n_s)$ shows a sharp peak at $n_s = s/2 = 20$. As $u$ increases, a new peak develops for $n_s=s$ and finally for very large $u$  all $n_s$ values have nonzero probability. This is consistent with the fact that for large $u$, the defect can be anywhere in the system, including empty regions (which is the most probable $n_s$ value in this case since we work with low particle density). 

\subsection{Long-range order in particle density and velocity}
\label{sec:dv}

\begin{figure}[H]
\centering
\includegraphics[scale=0.6]{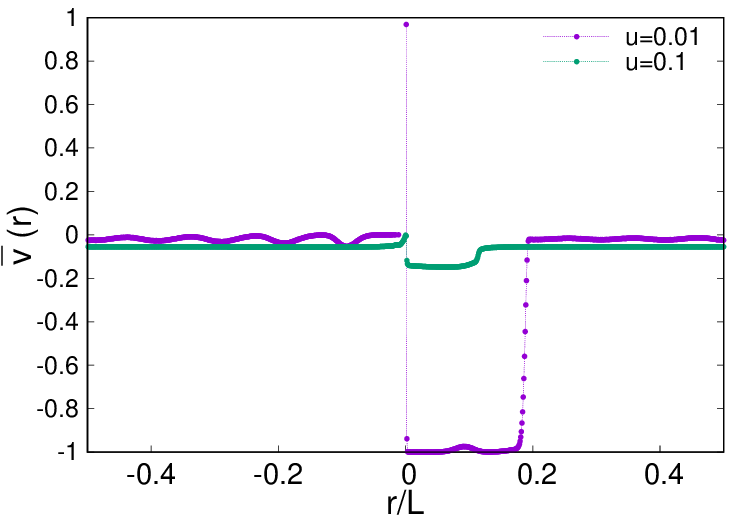}
\caption{Average velocity $\overline{v}(r)$ plotted as a function of $r/L$, where $r$ is the distance from the defect site. For small $u$, defect-induced long-range velocity ordering is evident from the presence of all left-movers except the one on the immediate right of the defect site. For large $u$ such ordering becomes weaker and the magnitude of $\overline{v}(r)$ decreases. We have used $\rho=0.2, L=1000$ here.} 
\label{fig:vx}
\end{figure}

In the previous two sections we have argued how the defect-induced velocity correlations give rise to cluster formation in steady state. To explicitly show the velocity ordering, we measure the velocity profile in the system. Let $\overline{v}(r)$ denote the average velocity at a distance $r$ from the defect site. A left-mover (right-mover) at this position contributes velocity $-1$ ($+1$) to the average, and if no particle is present, the contribution is zero. In Fig. \ref{fig:vx} we plot $\overline{v}(r)$ as a function of $r$ for two different defect velocities. Our data show that for both $u$, indeed there is a long stretch ahead of the defect where average velocity is negative. For small defect velocity, the defect is most likely to be found at the left edge of the particle cluster [also see data in Fig. \ref{Pls}(a)]. In this case, except at the site next to the defect, where a right-mover is found with high probability, almost all the particles in the cluster are left-movers and average velocity is very close to $-1$ over the stretch of the largest particle cluster. Note that a positive peak in front of the defect is present for all $u$. When a left-mover encounters the defect, it tumbles,  becomes a right-mover and moves ahead of the defect, which explains the presence of this peak. Behind the defect, it is mostly an empty region, but a few left-movers are present that escaped the cluster earlier and are passing through the empty region to rejoin the cluster at the right end. For larger $u$, the defect can sometimes move into the bulk of the cluster [as seen from Fig. \ref{Pls}(c)] and the velocity alignment is less pronounced but can still be observed to be present in this case. These measurements quantitatively show that the moving defect gives rise to long-range order in particle velocity. Similarly, by measuring the density profile as a function of distance from the defect site, one can observe long-range order in particle density as well (data shown in Fig. \ref{dpg0}). 

\begin{figure}[H]
\centering
\includegraphics[scale=1.1]{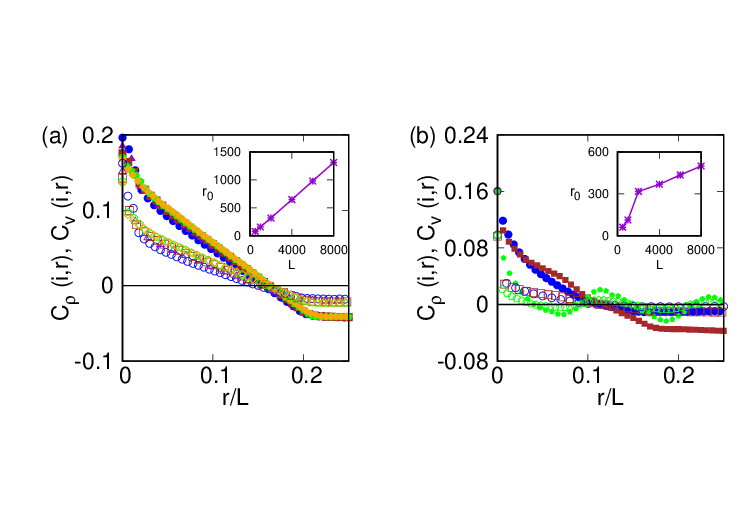}
\caption{Equal-time two-point correlation functions for densities and velocities are plotted for different $L$. Data shown for (a) $u=0.1$ and (b) $u=1$, with $L = 1000$ (blue), $2000$ (dark violet), $4000$ (brown), $6000$ (green), and $8000$ (orange). We have used closed solid symbols for density correlation and closed empty symbols for velocity correlation. Inset shows variation of the distance of zero crossing with system size.}  \label{dcrg0}
\end{figure}

To study the length scales associated with the density and velocity ordering, we measure two point correlation functions for these variables. Let $P(i, i+r)$ denote the joint probability to find occupied sites at distances $i$ and $(i+r)$ from the defect site. Then $C_\rho(i,r)$ is defined as  $C_\rho(i,r) = P(i, i+r) - \rho_i \rho_{i+r}$. Similarly, for velocity, we define $P_-(i, i+r) $ as the joint probability to find a left-mover at positions $i$ and $(i+r)$ from the defect site and  $C_v (i,r) = P_-(i, i+r) - P_-(i) P_-(i+r)$. In Fig. \ref{dcrg0} we plot $C_\rho(i,r)$ and $C_v (i,r)$ for $\rho=0.2$ and different $L$ values, as a function of $r/L$ for $i/L=0.1$. For small $u$ both the functions show scaling collapse. Moreover, $C_\rho(i,r)$ and $C_v (i,r)$ cross zero almost at the same distance $r_0$, which scales linearly with $L$ [Fig. \ref{dcrg0}(a), inset]. For large $u$, although linear scaling of $r_0$ with $L$ is observed for large $L$, the scaling collapse of $C_\rho(i,r)$ and $C_v (i,r)$ is not found. 

\section{Finite tumbling rate in the bulk: \boldmath $\gamma \neq 0$} \label{gn0}

\begin{figure}[h!]
\centering
\includegraphics[scale=1.5]{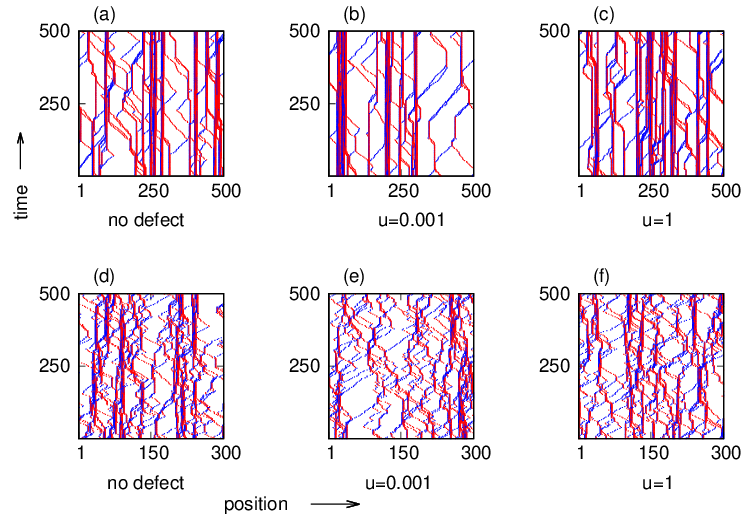}
\caption{Steady-state time-evolution plots for finite $\gamma$. A segment of the configuration is shown here. The top panel is for $\gamma=0.01$ and the bottom panel is for $\gamma=0.05$. Right-movers (left-movers) are marked by blue (red). }
\label{tmevgn0}
\end{figure}

For $\gamma \neq 0$, the active particles can flip their velocities even when the defect is not present. This allows the system to come out of the jammed configurations (Fig. \ref{fig:jam}).  In the absence of any external drive, earlier studies on systems of hardcore active particles \cite{slowman2016jamming, dandekar2020hard, soto2014run} have shown that in the long-time limit, density inhomogeneity develops in the system. The particles and holes partially phase separate and the steady state consists of multiple regions with densely packed particle clusters, separated by finite-sized hole clusters. Such phase separation can be observed even in the range of small $\rho$ values \cite{gutierrez2021collective,  barberis2019phase}. As $\gamma$ increases, the phase separation becomes weaker and the system approaches a disordered state. In Figs. \ref{tmevgn0}(a) and \ref{tmevgn0}(d) we have shown the time evolution of steady-state configurations for two values of $\gamma$, when there is no defect present in the system. In this section, we examine what happens when a periodically moving defect is introduced in the system. We are mainly interested in whether there is any particular range of defect velocity that might enhance the ordering in steady state. In Fig. \ref{tmevgn0} we show steady-state time evolution for two different defect velocities and compare it with the case of no defect. These plots indicate that a slowly moving defect may in fact aid in particle-hole phase separation. However, these time-evolution plots are rather qualitative and below we carry out more quantitative measurements to study the effect of varying defect velocity on the ordering. We perform our measurements for $\gamma=0.01$ and $0.05$ and throughout use $\rho=0.2, L=1000$ in this section. 

\subsection{Defect velocity affects density inhomogeneity} \label{lmaxgn0}

\begin{figure}[h!]
\centering
\includegraphics[scale=1.0]{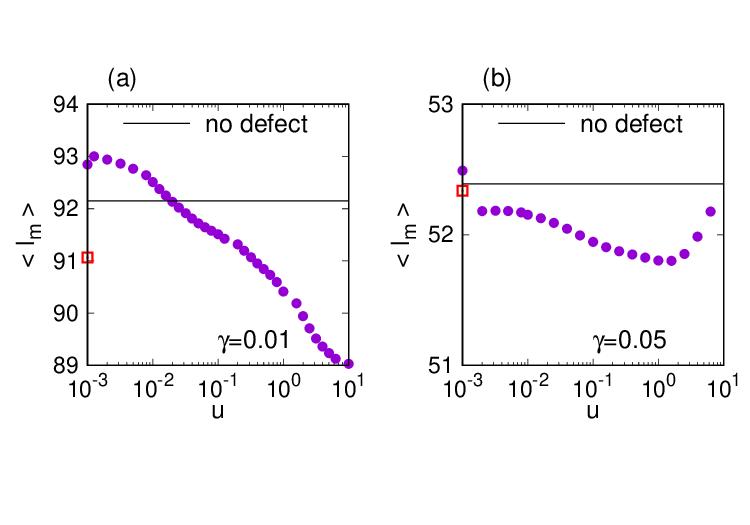}
\caption{Variation of average size of the largest hole cluster $\langle l_m \rangle$ with $u$. For a small $\gamma$, the defect causes a slight increase in $\langle l_m \rangle$ compared to the no defect case. The red square marks the point for $u=0$.}
\label{lmaxu_g01}
\end{figure}

Unlike what we have seen for $\gamma=0$, the moving defect does not give rise to a single large particle cluster or hole cluster for $\gamma=0.01$ or  $0.05$. Instead, there are multiple particle clusters and hole clusters present in the system whose sizes are finite and do not scale with $L$. To quantitatively study the order present in the system, we measure the average size of the largest hole cluster $\langle l_m \rangle$ for different values of $u$, as studied in the previous section. We present our data in Fig. \ref{lmaxu_g01}. For comparison, we also show the largest hole cluster in absence of defect, by a solid line. These data show that depending on $\gamma$, for small $u$, the presence of the moving defect may cause a slight increase in $\langle l_m \rangle$ value. For large $u$, the value of $\langle l_m \rangle$ falls below the solid line. In the same plots $\langle l_m \rangle$ for a static defect $u=0$ is also shown by a red square. Our data show that the $u$ dependence is rather weak. For $\gamma=0.01$, Fig. \ref{lmaxu_g01}(a) shows that $\langle l_m \rangle$ changes by four units, as $u$ varies over four orders of magnitude. For larger $\gamma$ the dependence is almost negligible, as seen from data in  Fig. \ref{lmaxu_g01}(b). We have also checked (data not shown here) that our conclusions remain unchanged when a higher particle density is considered. 

\begin{figure}[h!]
\centering
\includegraphics[scale=1.0]{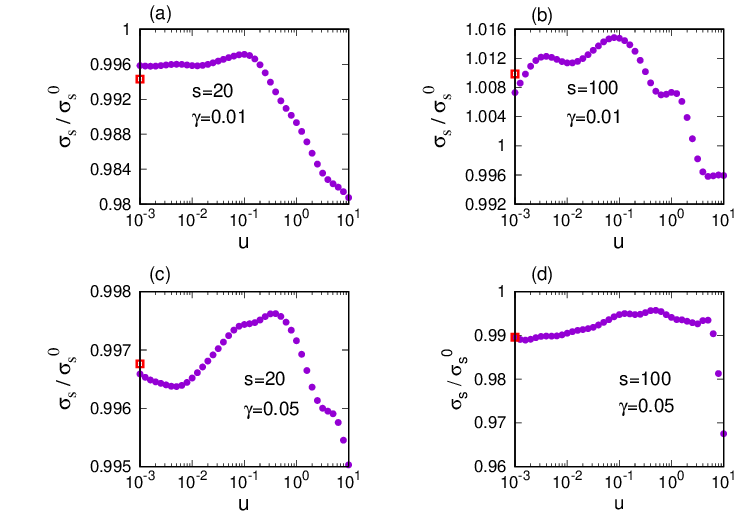}
\caption{Subsystem number fluctuations for holes as a function of $u$. The red square represents the $u=0$ point. Small and moderate $u$ values slightly enhance the density fluctuations for a large subsystem size. }
\label{sigmaun0}
\end{figure}

Another way to quantitatively characterize the density inhomogeneity in the system is by measuring subsystem density fluctuations. Consider a subsystem consisting of $s$ consecutive sites, with $s \ll L$. We measure the density fluctuations in this subsystem by calculating the standard deviation $\sigma_s$ of the number of holes present in it. We compare $\sigma_s$ with a similar quantity $\sigma_s^0$ measured in absence of a defect. In Fig. \ref{sigmaun0} we present our data. The plots show that depending on the choice of $s$, the density inhomogeneity may get enhanced or suppressed by the defect. In Fig. \ref{sigmaun0} we have used two values of $s$, which correspond to two different length scales present in the system. For $\gamma=0.01$ the average size of a hole cluster is $20$ (data not shown) and the largest hole cluster size is $\simeq 90$. Our choice of $s$ in Fig. \ref{sigmaun0} is based on this. Our data in Figs. \ref{sigmaun0}(a) and \ref{sigmaun0}(b)  show that the defect suppresses the density inhomogeneity at a smaller length scale ($s=20$) while for a larger length scale ($s=100$) the ratio $\sigma_s / \sigma_s^0$ is slightly larger than $1$ for small $u$, indicating enhanced inhomogeneity in density. 

To explain this interesting effect we note that the defect plays two kinds of roles in our system. On one hand, it increases the effective rate of velocity reversal, which allows the particles to escape from a locally jammed configuration faster. On the other hand, the moving defect gives rise to velocity ordering around its position, which in turn gives rise to ordering in density. It is this second effect which facilitates the formation of particularly large clusters of particles and holes in the system. In the case of a small $\gamma$, this effect is stronger for small $u$ and the size of the largest cluster (both particles and holes) is bigger compared to the no-defect case. This is clearly shown in our data in Fig. \ref{lmaxu_g01}(a). Our observation of $\sigma_s > \sigma_s^0$ for $s=100$ in Fig. \ref{sigmaun0}(b) is consistent with it. As $u$ increases,  although velocity ordering is still there, the particles do not get enough time to form stable clusters before the defect comes back to break them up. In this range of $u$, the particles tend to have a higher mobility (supporting data in Fig. \ref{J_u_g0p01}), which tends to make the system more homogeneous and  $\sigma_s$ drops. However, the smaller clusters in the system are mainly formed because of local jamming between particles with opposite velocities. The frequent velocity reversal caused by the moving defect allows the particles to escape the clusters and move into the empty region in between. This mixing makes the system more homogeneous and this is why in Fig. \ref{sigmaun0}(a) $\sigma_s$ is consistently less than $\sigma_s^0$ for all values of $u$, when $s$ is small.

For higher $\gamma$, our data in Figs. \ref{sigmaun0}(c) and \ref{sigmaun0}(d) show that $\sigma_s$ remains smaller than $\sigma_s^0$ for both small and large $s$. We again use $s=20,100$ here, to be able to compare with the smaller $\gamma$ case. Since the system becomes homogeneous with increasing $\gamma$, the above $s$ values are larger than mean cluster size and largest cluster size, respectively. Over these length scales, introduction of the moving defect does not enhance the density inhomogeneity; rather it destabilizes the cluster edges and pushes the system toward a more mixed  state. When $u$ is large, $\sigma_s/\sigma_s^0$ sharply decreases with $u$, for all $s$ and $\gamma$. In this case the frequent velocity reversal caused by the defect destroys the velocity ordering in the system (also see Sec. \ref{sub:dv} below) and density inhomogeneity is lost.

In Fig. \ref{HMgn0} we present a heat map for $\sigma_s/ \langle n_s \rangle$ in the $\gamma$-$u$ plane for $s=100$. Here $\langle n_s \rangle = (1-\rho)s$ stands for the average number of holes in a segment of size $s$. This heat map shows largest values of $\sigma_s$ is recorded near the bottom left corner of the graph, which means for small $\gamma$ and small $u$ the density inhomogeneities are most pronounced. For small $\gamma$, as $u$ increases, $\sigma_s$ decreases, but for large $\gamma$ the $u$ dependence almost disappears. This is expected since defect movement does not affect the system much when the bulk tumbling rate is high. 
\begin{figure}[h!]
\centering
\includegraphics[scale=1.0]{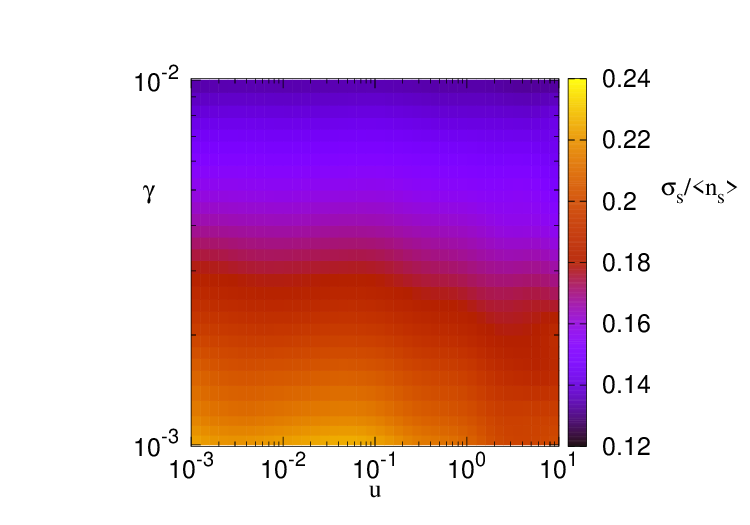}
\caption{Fluctuation $\sigma_s$ in hole numbers for a subsystem of size $s$, scaled by average hole number $\langle n_s \rangle$, plotted in the $\gamma$-$u$ plane. For small $\gamma$ and small $u$ fluctuations are most significant. } \label{HMgn0}
\end{figure}

\subsection{Density and velocity profile around the defect} \label{sub:dv}

\begin{figure}[h!]
\centering
\includegraphics[scale=1.0]{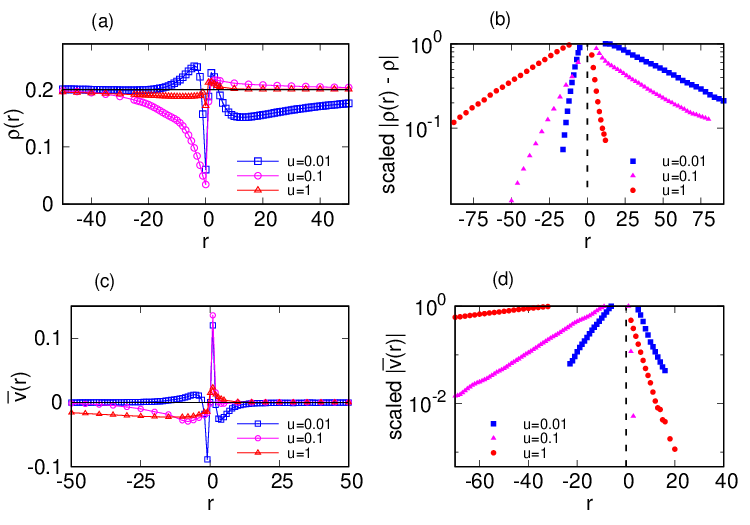}
\caption{Density and velocity profiles around the defect site for different $u$ values. As a function of distance $r$ from the defect site, both $\rho(r)$ and $\overline{v}(r)$ show exponential variation. We have used $\gamma=0.01$ here.}
\label{fig:dpvp}
\end{figure}

The moving defect creates a density profile and a velocity profile around its position. In Fig. \ref{fig:dpvp} we plot these profiles as a function of distance from the defect site for few different $u$ values. We find both density and velocity decay exponentially with length scales depending on $u$. Our data in Figs. \ref{fig:dpvp}(b) and \ref{fig:dpvp}(d) also show that the length scales are very different on two sides of the defect. While the velocity profile always decays fast in front of the defect and more slowly behind it, the density profile shows a $u$-dependent asymmetry. For small $u$, the density decays sharply behind the defect but over a longer length scale in front of the defect, while for large $u$ the trend is reversed. 

Since the defect reverses the velocity of the particles, at the defect site the average velocity is always zero. For $\gamma=0$ we had argued in Sec. \ref{sec:g0t} that behind the defect there is a trail of left-movers. However, for finite $\gamma$ whether such a trail exists depends on the competition between $u$ and $\gamma$. While the defect site still tends to create such a trail behind it, due to finite $\gamma$ the velocities are randomized again before the particles are able to travel long. For small $u$, in our stroboscopic measurement, we indeed find $\overline{v}(r)$ decaying over a short length scale [Fig. \ref{fig:dpvp}(d)] behind the defect. However, as $u$ increases, there is not enough time for randomization of $v$ before the defect moves to a new site. In this case, over a longer spatial range we find negative values of $\overline{v}(r)$. In front of the defect the decay of $\overline{v}(r)$ is sharper, especially for moderate $u$ values. Our data in Fig. \ref{fig:dpvp}(d) show a nonmonotonicity in decay length scale with the $u$, which we do not understand. In Fig. \ref{fig:vp_te} in Appendix \ref{A5} we show the time evolution of the velocity profile and compare it with the $\gamma =0$ case.

The behavior of density profiles can be understood from $\overline{v}(r)$ vs $r$ plots. For small $u$,  during one residence time of the defect,  the left-movers which have left the defect site and moved behind it, have their velocities randomized owing to a finite $\gamma$ and a local jammed state is created behind the defect, which causes a density peak there. However, as $u$ increases, the residence time of the defect becomes smaller compared to $1/\gamma$, and during this time the left-movers behind the defect have freely moved away. Thus, a density trough is created behind the defect, which relaxes over a long distance to the bulk density.

\section{Large persistence time helps the defect restore long-range order } \label{orderdisorder}

In the previous two sections we have seen how a slowly moving defect gives rise to a single large cluster and how the presence of bulk tumbling breaks it apart. The formation of a large cluster is a consequence of long-range order in particle velocity which the moving defect creates along its trajectory. A finite $\gamma$ allows the particles to switch their velocity independently, irrespective of the defect position. This destroys the velocity ordering in the system and the large cluster cannot be sustained anymore. However, if $\gamma$ is small enough such that the time interval between two tumbling events is larger than or comparable to the time taken by the defect to make one complete sweep of the entire ring lattice, then it is possible that the loss in velocity ordering due to spontaneous tumbling is restored by the defect. In other words, during the time interval $u/L$ if a relatively small number of spontaneous tumbles occur in the bulk, then the loss in velocity alignment, and consequently the loss in density ordering, happens slowly such that the moving defect can come back during the next sweep and reinstate the order. In this limit, one expects long-range order in both density and velocity, as seen for the $\gamma = 0$ case. In fact, by tuning the $\gamma$ value in this range it should be possible to see how ordering slowly gets destroyed. In this section we examine this scenario.

\begin{figure}[h!]
\centering
	\vspace{-2. cm}
\includegraphics[scale=1.4]{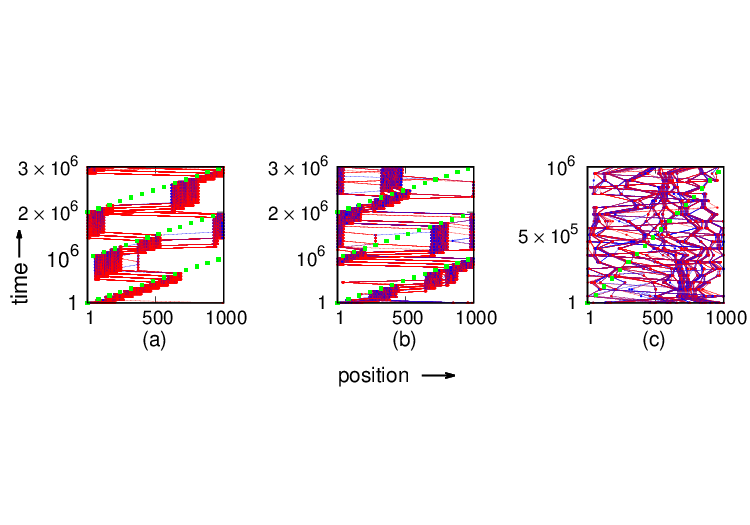}
	\vspace{-2. cm}
\caption{Steady-state time-evolution plots for (a) $\gamma = u/L$, (b) $\gamma = 10u/L$, and (c) $\gamma = 1000u/L$. We have used $u=0.001$ and $L=1000$ here. Right-movers (left-movers) are marked by blue (red). The position of the defect is marked by green dots. These data are for $\rho = 0.2$.} 	\label{tmevgsmall}
\end{figure}

Note that the choice of $u$ is important here. If $u$ is small, then the particles form a single cluster with almost perfectly aligned velocities for $\gamma =0$. For $\gamma \lesssim u/L$ the presence of bulk tumbling affects this order only slightly which is restored when the defect revisits these locations. In fact we find that even when $\gamma$ exceeds $u/L$ the ordering is not immediately lost, although it weakens. On the other hand, when $u$ is moderately large, then even for $\gamma =0$ we do not have perfect order. In such a case, ordering is lost soon after $\gamma$ crosses $ u/L$. In Fig. \ref{tmevgsmall} we show the time evolution for $\gamma = u/L, 10u/L, 1000u/L$ for a small value of $u$. For $\gamma = u/L$ our plot in Fig. \ref{tmevgsmall}(a) shows that a single large particle cluster survives in this case and long-range order in velocity is also recovered. However, comparison with Fig. \ref{fig:g0t} also shows that the velocity ordering is not as perfect as seen for $\gamma =0$ case. Spontaneous tumbling happening in the bulk does affect the velocity alignment to some extent. For $\gamma = 10 u /L$ velocity ordering is even less. Moreover, in this case one large cluster is often unstable and breaks into two or more large clusters. Unlike what we had seen for $\gamma =0$, where particles detach from one end of the cluster to join the other end, in this case there is a significant probability that the particles tumble before joining the cluster and form a local jammed state. As more particles leave the main cluster and get stuck in this jam, a new cluster emerges in this region which is away from the defect site and away from the other (main) cluster. Figure Fig. \ref{tmevgsmall}(c) shows the configurations for $\gamma = 1000 u/L$. In this case large clusters are not seen anymore and the behavior of the system is similar to what we have reported in Sec. \ref{gn0}.

\begin{figure}[h!]
\centering
\includegraphics[scale=0.6]{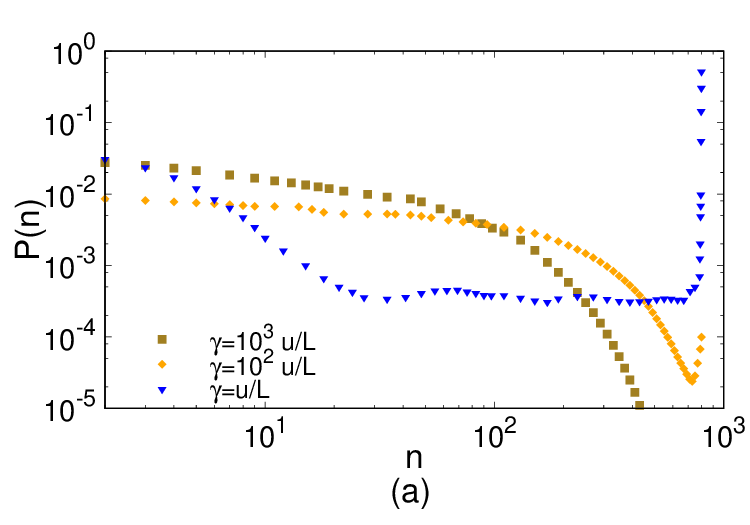}
\includegraphics[scale=0.6]{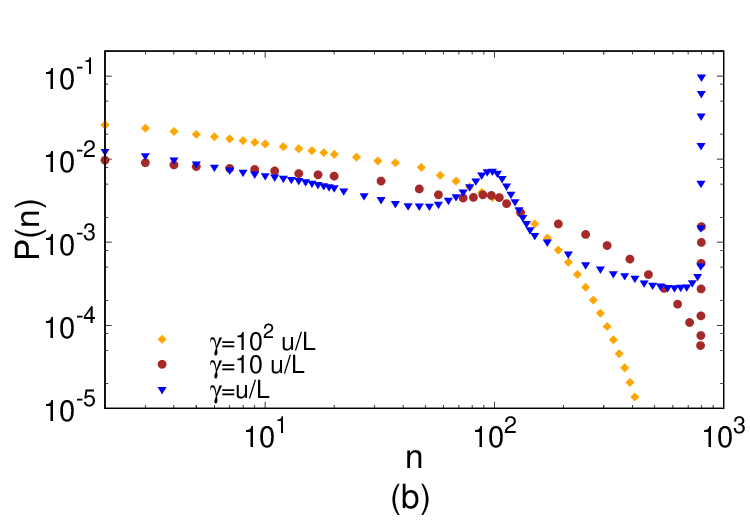}
\caption{Probability distribution $P(n)$ for hole clusters of size $n$. Data for (a) $u=0.001$, and (b) $u=0.01$. We have used $L=1000$ and $\rho=0.2$ here.}
\label{csdsmallgammar0p2} 
\end{figure}

\begin{figure}[h!]
	\centering
	\includegraphics[scale=1.5]{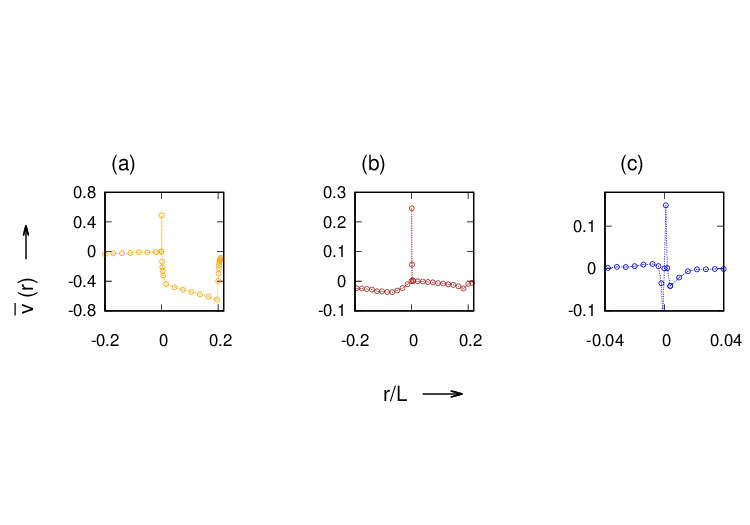}
	\caption{Velocity profiles $\overline{v}(r)$ as a function of scaled distance $r/L$ from the defect site. (a) $\gamma = u/L$, (b) $\gamma = 10u/L$, and (c) $\gamma = 1000 u/L$. We have used $u=0.001$ and $L=1000$ here.}
	\label{fig:vplowgamma}
\end{figure}

\begin{figure}[h!]
\centering
\includegraphics[scale=0.6]{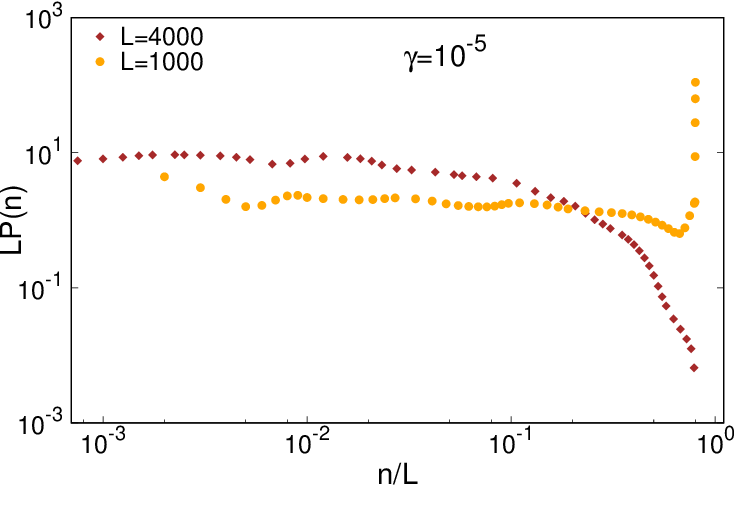}
\caption{For a fixed $\gamma$ and keeping all other parameters the same, the size distribution $P(n)$ of hole clusters shows a peak for $L=1000$ but for $L=4000$ the peak disappears.}
	\label{fig:csdh_L}
\end{figure}

To further characterize the order present in the system, we plot in Fig. \ref{csdsmallgammar0p2} the hole cluster distribution for small and moderate $u$ values. For small $u$, Fig. \ref{csdsmallgammar0p2}(a) shows that even when $\gamma = 100 u /L$ there is a peak at large $n$, indicating the presence of large clusters. But for moderate $u$ [Fig. \ref{csdsmallgammar0p2}(b)] no peak is observed in $P(n)$ when $\gamma = 100 u /L$. Figure \ref{csdsmallgammar0p2}(b) also shows that for $\gamma = u/L$ instead of a single peak at large $n$, an additional peak develops for intermediate $n$, indicating the presence of multiple clusters. This is consistent with the mechanism discussed in the previous paragraph. In Fig. \ref{fig:vplowgamma}, we examine how velocity ordering decreases with $\gamma$. We plot average velocity as a function of distance from the defect site. For $\gamma = u/L$ the velocity profile confirms a long-range velocity alignment with most particles as left-movers [Fig. \ref{fig:vplowgamma}(a)]. For  $\gamma = 10 u/L$ [Fig. \ref{fig:vplowgamma}(b)] velocity ordering is much weaker and when $\gamma = 1000 u/L$ [Fig. \ref{fig:vplowgamma}(c)], the velocity profile shows only short-range variation, as seen earlier in Fig. \ref{fig:dpvp}(c). In Fig. \ref{fig:csdh_L} we check the system size dependence by plotting $P(n)$ vs $n$ for two different $L$ values and fixed $\gamma$. The moving defect takes longer to sweep the full system when $L$ is increased, and a larger number of tumbles which take place at the bulk during this time interval makes it more difficult for the defect to restore order. Consistent with this, we show in Fig. \ref{fig:csdh_L} that there is no peak in $P(n)$ for larger $L$. Throughout this section, we have considered $\rho = 0.2$ and we have also checked (data not shown) that choosing a different $\rho$ does not affect the conclusions. 

\section{Conclusions} \label{sec:con}

In this paper, we studied a persistent exclusion process in the presence of a moving defect. The model consists of a set of hard-core particles on a one-dimensional ring lattice; each particle has a velocity (positive or negative) associated with it. The particles perform stochastic jumps in the direction of the velocity, while satisfying the hard-core exclusion. The velocity of each particle can flip with a small probability $\gamma$. The defect is introduced in the form of a special site where the particles tumble with probability $1$. This defect site moves around the lattice with speed $u$. We are interested in how the introduction of the moving defect affects the steady-state properties of the system.

Our Monte Carlo simulations show that the moving defect has a dramatic effect on the system for $\gamma=0$. While the system gets stuck in a jammed configuration with no long-range order in the absence of a defect, it goes to a strongly phase separated state when the defect is present. For small values of $u$ almost all particles in the system form a single large cluster. As $u$ is increased, the single cluster breaks into multiple smaller clusters. We show that the phase separation is a direct consequence of the defect-induced long-range order in particle velocity. For small $u$ we measure the two point density-density and velocity-velocity correlation functions and find $r/L$ scaling, as expected in a phase-ordered system. For nonzero $\gamma$ a competition ensues between the timescales associated with bulk tumbling and the defect movement. While the periodically moving defect attempts to create long-range order in the system, spontaneous tumbling at random positions tends to destroy the order.  However, if $\gamma$ is small enough then the loss of order due to tumbling happens slowly such that there is sufficient time for the moving defect to restore the order. In that case, the system can still support a single large cluster and long-range velocity correlations. As $\gamma$ increases, the ordering gradually goes away. The time required for the defect to move through the entire ring lattice and come back to the same position is $L/u$. If too many tumbles take place during this time, the defect can not restore the long-range order anymore and the system breaks into multiple high-density and low-density patches. The competition between the two timescales is manifested in the density profile and velocity profile around the defect position.

It should be possible to test our conclusions in experiments. {\sl E.coli} bacteria are one of the most commonly encountered natural examples of active particles with run-and-tumble motility. The presence of a chemical gradient in the medium causes the cells to change their tumbling rate and perform chemotaxis. In many experiments \cite{berg1990chemotaxis, binz2010motility, li2017barrier} bacterial motility is studied inside microfluidic channels whose width is comparable to (or shorter than) the distance traveled by a cell during a run. In such situations the motion of the bacterial cells can be considered effectively one dimensional. It has even been possible to isolate certain mutant strains of {\sl E.coli} whose tumbling ability has been impaired \cite{silverman1976identification, parkinson1976chea, parkinson1982isolation, parkinson1979interaction, sanna1996vivo}. Such cells can only run and are unable to tumble, just like what we have considered in Sec \ref{g0}. The defect can be  implemented by a potential barrier, in the form of a hard wall, which the cells cannot cross: when they hit the wall, they reverse their velocity. The defect movement can be mimicked by removing the partition at the end of its residence time and immediately reinserting it at a small distance away. In other words, a narrow microfluidic channel in the shape of a ring, with removable partitions and motile {\sl E.coli} bacteria inside it can serve as a model system to test our conclusions experimentally.

To the best of our knowledge, this is the first study of an active matter system in a time-periodic external drive. Here we have considered a simple model of hard-core run-and-tumble particles to describe active matter. But our study can be generalized for other active matter systems such as active Brownian particles with attractive and aligning interactions, or the Vicsek model, or any other model systems which are used to study active matter. What happens to such systems when they are subjected to a time-periodic potential and the effect of varying the periodicity or strength of that potential are interesting open questions which are expected to unravel rich physics.

\section{Acknowledgement}
SC acknowledges support from Anusandhan National Research Foundation (ANRF), India (Grant No. CRG/2023/000159).

\appendix

\section{Size distribution of particle clusters for \boldmath$\gamma=0$}{\label{A1}}

\begin{figure}[h!]
\centering
\includegraphics[scale=0.5]{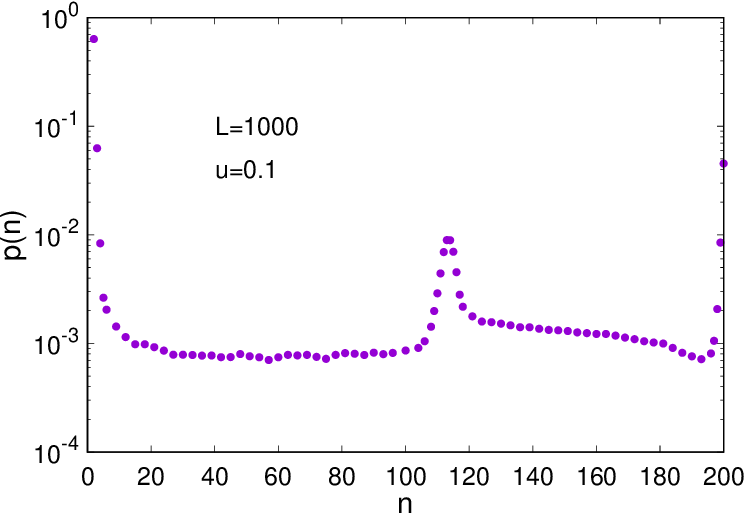}
\caption{Probability distribution $p(n)$ for size of the particle clusters for $\gamma=0$ and $u=0.1$. It shows that the largest peak is formed at $n=N$. The intermediate peak at $n < N$ is an outcome of one particular mechanism for defect-induced evolution of the largest particle cluster, explained in detail in Sec. \ref{csdg0} of the main text. }
\label{csdpg0u0p1}
\end{figure}

In Fig. \ref{csdpg0u0p1} we show the data for the probability distribution of the size of the particle clusters for a fixed $u$ when $\gamma =0$. The distribution shows a large peak at $n=N$, consistent with a complete phase-separated state. Such a peak at the largest cluster is also seen in the distribution of hole clusters in Fig. \ref{csd0}(a). However, unlike the hole cluster data, here we observe a secondary peak at $n < N$. This intermediate peak arises as a consequence of the fragmentation of the largest particle cluster when the defect passes through it (representative configuration shown in Fig. \ref{fig:fragment}). The nonmonotonic variation of $\langle l_m \rangle$ with $u$ as shown in Fig. \ref{lmaxu} is also an outcome of this effect.

\section{Density profile about the defect site for $\gamma=0$}
\begin{figure}[h!]
\centering
\includegraphics[scale=0.5]{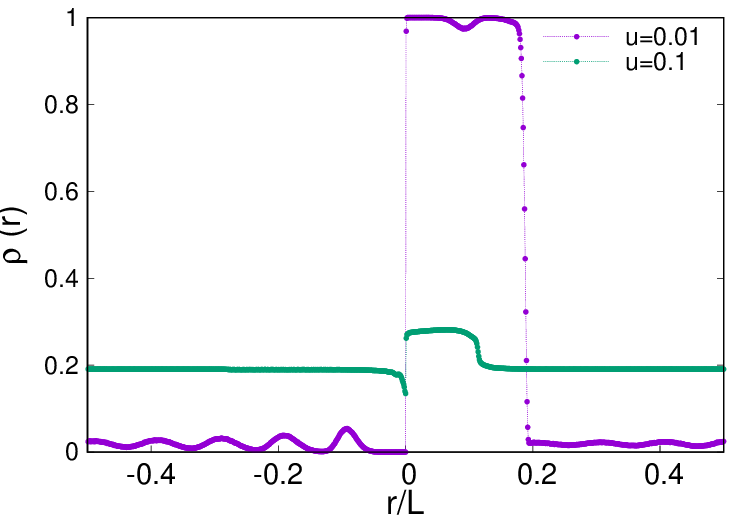}
\caption{Density profile for particles as a function of distance from the defect site with $u=0.01, 0.1$. The distance $r$ has been scaled by the system size $L=1000$. $\rho=0.2$ here.}  \label{dpg0}
\end{figure}

{\label{A2}}
We show the spatial ordering in particle density as mentioned in Sec. \ref{sec:dv} by measuring the average density as a function of distance from the defect site [Fig. \ref{dpg0}]. For small $u$, the average density in front of the defect is $1$ up to $r/L \simeq \rho$, which clearly shows that the system supports the presence of a single large cluster with almost all the particles, and the defect present at its left edge. Small periodic peaks behind the defect correspond to the presence of particles earlier detached from the cluster that are now moving leftward. For moderate $u$, the defect is not always at the left edge; it can move inside the cluster and often the cluster can be fragmented as shown in Fig. \ref{fig:fragment}. For both $u$, the density profiles are consistent with the velocity profiles presented in Fig. \ref{fig:vx} of the main text. This shows that long-range order in the density of the system is a direct consequence of defect-induced velocity ordering of the particles.

\section{Variation of $\langle l_m \rangle$ with $u$ for different particle densities}
\label{app:lm}

\begin{figure}[h!]
\centering
\includegraphics[scale=1]{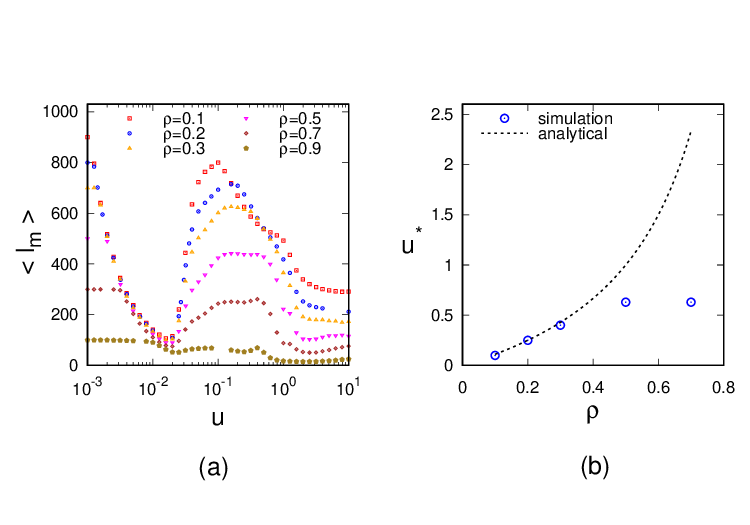}
\caption{(a) $\langle l_m \rangle$ vs $u$ data for different $\rho$, where $L=1000$. In the regime of small $u$, the variation is independent of $\rho$, while the behavior changes slightly in the intermediate-$u$ regime as $\rho$ is varied. The distinct peak at $u^*$ in $\langle l_m \rangle$ observed for a small $\rho$ turns into a plateau as $\rho$ increases. A subsequent fall in $\langle l_m \rangle$ starts at higher $u=u^*$, for larger density. (b) Variation of $u^*$ with $\rho$ showing our analytical estimation does not hold beyond small values of $\rho$, where predicted $u^*$ becomes large. } 
\label{fig:lmax_u_rho}
\end{figure}

In this section we present the variation of largest hole cluster size $\langle l_m \rangle$ with defect velocity $u$ for different values of particle density $\rho$. Like Fig. \ref{lmaxu} these data are for $\gamma =0$. We find similar qualitative behavior for different $\rho$, except when $\rho$ is very large [Fig. \ref{fig:lmax_u_rho}(a)].  From these data we read off the defect velocity $u^\ast$ at which $\langle l_m \rangle$ starts decreasing from the peak it reaches at intermediate $u$ values. In Fig. \ref{fig:lmax_u_rho}(b) we plot these $u^\ast$ values with $\rho$ and compare them with our analytical estimate $\rho/(1-\rho)$. For small $\rho$ we find agreement but for large $\rho$ there is significant overestimation. Our analytical calculation was based on the simple assumption that as long as $u \lesssim u^\ast$ the system can support a single large cluster. However, this assumption breaks down for large values of $u$ when a fast-moving defect gives rise to weak velocity ordering in the system and large clusters cannot form anyway. Our data in Fig. \ref{fig:lmax_u_rho}(a) also show different qualitative behavior seen for very large $\rho$, for which we do not have any simple explanation. 

\section{Variation of particle current with defect velocity}  \label{A4}

\begin{figure}[h!]
\centering
\includegraphics[scale=0.5]{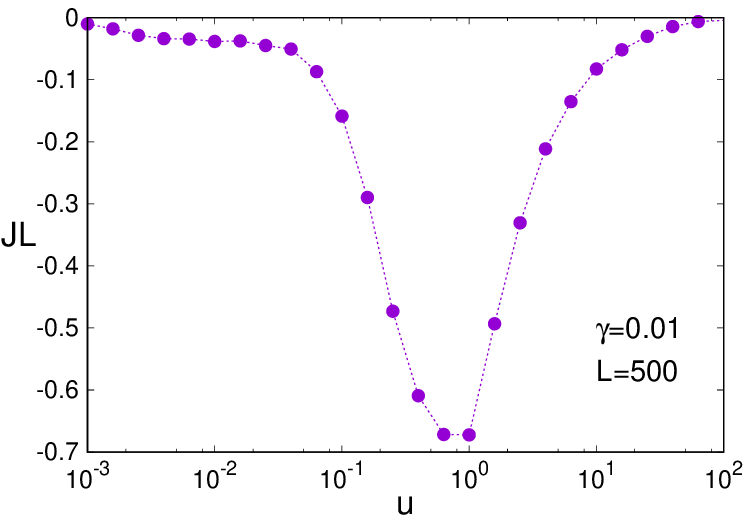}
\caption{Scaled current $JL$ is plotted against $u$ for $\gamma=0.01$, $L=500$. The variation is nonmonotonic showing a large peak of negative current at a moderate $u$ value. } \label{J_u_g0p01}
\end{figure}

In Fig. \ref{J_u_g0p01} we plot the scaled current as a function of defect velocity. Due to the directional motion of the defect, a current $J$ is generated in the system. Since there is only one defect site, $J \sim 1/L$ scaling is expected \cite{chatterjee2014interacting,  chatterjee2016symmetric, das2023optimum}. The scaled current remains negative and has a large magnitude for moderate $u$ values. This observation is consistent with our explanation in Sec. \ref{lmaxgn0}. 

\section{Time evolution of velocity profiles towards steady state}  \label{A5}

\begin{figure}[h!]
\centering
\includegraphics[scale=1.1]{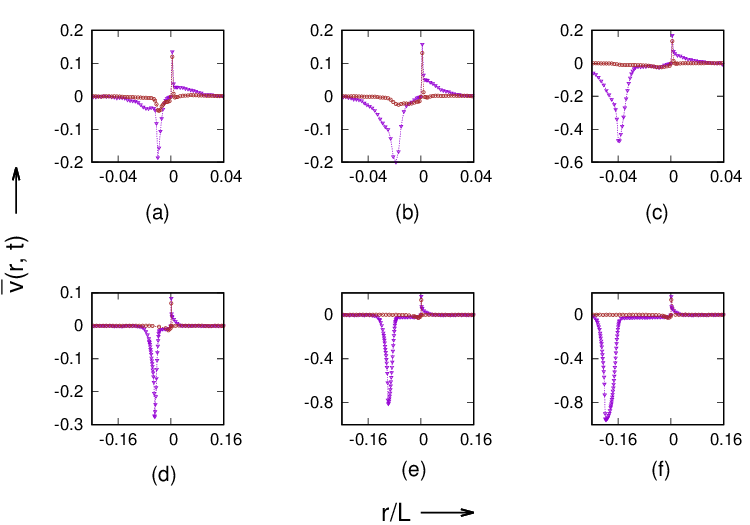}
\caption{Average velocity  $\bar{v} (r,t )$ shown as a function of the distance from the defect for different times as the system evolves toward steady state from a random initial configuration. Data presented for $\gamma=0$ (purple) and $\gamma=0.01$ (brown) at times (a) $100$, (b) $200$, (c) $400$, (d) $500$, (e) $1000$, and (f) $2000$. Here $u=0.1$, $L=1000$ and $\rho=0.2$.} \label{fig:vp_te}
\end{figure}

To get further insights into the difference in steady-state velocity profiles for $\gamma =0$ and $\gamma = 0.01$ [Figs. \ref{fig:vx} and \ref{fig:dpvp}(c), respectively] we compare the time evolution of these velocity profiles for these two cases. Starting from a random initial configuration, we measure $\bar {v} (r, t)$ and plot as a function of $r$ for few different $t$ values in Fig. \ref{fig:vp_te}. Our data show that a positive peak in front of the defect and a negative peak behind it develop within short times [Fig. \ref{fig:vp_te}(a)]. While the magnitude of the positive peak is comparable for both $\gamma$ values, the negative peak gets much larger for $\gamma =0$ as time goes on. This negative peak for $\gamma =0$ also moves leftward, away from the defect site, until it moves through the entire periodic system and reaches on the right of the defect site, next to the positive peak, as found in Fig. \ref{fig:vx}. This evolution takes place over several time periods of defect movement (here data are shown up to a small fraction of one time period $L/u$). 

\begin{thebibliography}{48}
\expandafter\ifx\csname natexlab\endcsname\relax\def\natexlab#1{#1}\fi
\expandafter\ifx\csname bibnamefont\endcsname\relax
  \def\bibnamefont#1{#1}\fi
\expandafter\ifx\csname bibfnamefont\endcsname\relax
  \def\bibfnamefont#1{#1}\fi
\expandafter\ifx\csname citenamefont\endcsname\relax
  \def\citenamefont#1{#1}\fi
\expandafter\ifx\csname url\endcsname\relax
  \def\url#1{\texttt{#1}}\fi
\expandafter\ifx\csname urlprefix\endcsname\relax\def\urlprefix{URL }\fi
\providecommand{\bibinfo}[2]{#2}
\providecommand{\eprint}[2][]{\url{#2}}

\bibitem[{\citenamefont{Ramaswamy}(2017)}]{Ramaswamy_2017}
\bibinfo{author}{\bibfnamefont{S.}~\bibnamefont{Ramaswamy}},
  \bibinfo{journal}{Journal of Statistical Mechanics: Theory and Experiment}
  \textbf{\bibinfo{volume}{2017}}, \bibinfo{pages}{054002}
  (\bibinfo{year}{2017}).

\bibitem[{\citenamefont{Schweitzer and Farmer}(2003)}]{schweitzer2003brownian}
\bibinfo{author}{\bibfnamefont{F.}~\bibnamefont{Schweitzer}} \bibnamefont{and}
  \bibinfo{author}{\bibfnamefont{J.~D.} \bibnamefont{Farmer}},
  \emph{\bibinfo{title}{Brownian agents and active particles: collective
  dynamics in the natural and social sciences}}, vol.~\bibinfo{volume}{1}
  (\bibinfo{publisher}{Springer}, \bibinfo{year}{2003}).

\bibitem[{\citenamefont{Bechinger et~al.}(2016)\citenamefont{Bechinger,
  Di~Leonardo, L{\"o}wen, Reichhardt, Volpe, and Volpe}}]{bechinger2016active}
\bibinfo{author}{\bibfnamefont{C.}~\bibnamefont{Bechinger}},
  \bibinfo{author}{\bibfnamefont{R.}~\bibnamefont{Di~Leonardo}},
  \bibinfo{author}{\bibfnamefont{H.}~\bibnamefont{L{\"o}wen}},
  \bibinfo{author}{\bibfnamefont{C.}~\bibnamefont{Reichhardt}},
  \bibinfo{author}{\bibfnamefont{G.}~\bibnamefont{Volpe}}, \bibnamefont{and}
  \bibinfo{author}{\bibfnamefont{G.}~\bibnamefont{Volpe}},
  \bibinfo{journal}{Reviews of modern physics} \textbf{\bibinfo{volume}{88}},
  \bibinfo{pages}{045006} (\bibinfo{year}{2016}).

\bibitem[{\citenamefont{Gr{\'e}goire and Chat{\'e}}(2004)}]{gregoire2004onset}
\bibinfo{author}{\bibfnamefont{G.}~\bibnamefont{Gr{\'e}goire}}
  \bibnamefont{and}
  \bibinfo{author}{\bibfnamefont{H.}~\bibnamefont{Chat{\'e}}},
  \bibinfo{journal}{Physical review letters} \textbf{\bibinfo{volume}{92}},
  \bibinfo{pages}{025702} (\bibinfo{year}{2004}).

\bibitem[{\citenamefont{Toner and Tu}(1995)}]{toner1995long}
\bibinfo{author}{\bibfnamefont{J.}~\bibnamefont{Toner}} \bibnamefont{and}
  \bibinfo{author}{\bibfnamefont{Y.}~\bibnamefont{Tu}},
  \bibinfo{journal}{Physical review letters} \textbf{\bibinfo{volume}{75}},
  \bibinfo{pages}{4326} (\bibinfo{year}{1995}).

\bibitem[{\citenamefont{Vicsek and Zafeiris}(2012)}]{vicsek2012collective}
\bibinfo{author}{\bibfnamefont{T.}~\bibnamefont{Vicsek}} \bibnamefont{and}
  \bibinfo{author}{\bibfnamefont{A.}~\bibnamefont{Zafeiris}},
  \bibinfo{journal}{Physics reports} \textbf{\bibinfo{volume}{517}},
  \bibinfo{pages}{71} (\bibinfo{year}{2012}).

\bibitem[{\citenamefont{Toner and Tu}(1998)}]{toner1998flocks}
\bibinfo{author}{\bibfnamefont{J.}~\bibnamefont{Toner}} \bibnamefont{and}
  \bibinfo{author}{\bibfnamefont{Y.}~\bibnamefont{Tu}},
  \bibinfo{journal}{Physical review E} \textbf{\bibinfo{volume}{58}},
  \bibinfo{pages}{4828} (\bibinfo{year}{1998}).

\bibitem[{\citenamefont{Tu et~al.}(1998)\citenamefont{Tu, Toner, and
  Ulm}}]{tu1998sound}
\bibinfo{author}{\bibfnamefont{Y.}~\bibnamefont{Tu}},
  \bibinfo{author}{\bibfnamefont{J.}~\bibnamefont{Toner}}, \bibnamefont{and}
  \bibinfo{author}{\bibfnamefont{M.}~\bibnamefont{Ulm}},
  \bibinfo{journal}{Physical review letters} \textbf{\bibinfo{volume}{80}},
  \bibinfo{pages}{4819} (\bibinfo{year}{1998}).

\bibitem[{\citenamefont{Vicsek et~al.}(1995)\citenamefont{Vicsek, Czir{\'o}k,
  Ben-Jacob, Cohen, and Shochet}}]{vicsek1995novel}
\bibinfo{author}{\bibfnamefont{T.}~\bibnamefont{Vicsek}},
  \bibinfo{author}{\bibfnamefont{A.}~\bibnamefont{Czir{\'o}k}},
  \bibinfo{author}{\bibfnamefont{E.}~\bibnamefont{Ben-Jacob}},
  \bibinfo{author}{\bibfnamefont{I.}~\bibnamefont{Cohen}}, \bibnamefont{and}
  \bibinfo{author}{\bibfnamefont{O.}~\bibnamefont{Shochet}},
  \bibinfo{journal}{Physical review letters} \textbf{\bibinfo{volume}{75}},
  \bibinfo{pages}{1226} (\bibinfo{year}{1995}).

\bibitem[{\citenamefont{Ginelli}(2016)}]{ginelli2016physics}
\bibinfo{author}{\bibfnamefont{F.}~\bibnamefont{Ginelli}},
  \bibinfo{journal}{The European Physical Journal Special Topics}
  \textbf{\bibinfo{volume}{225}}, \bibinfo{pages}{2099} (\bibinfo{year}{2016}).

\bibitem[{\citenamefont{Fily and Marchetti}(2012)}]{fily2012athermal}
\bibinfo{author}{\bibfnamefont{Y.}~\bibnamefont{Fily}} \bibnamefont{and}
  \bibinfo{author}{\bibfnamefont{M.~C.} \bibnamefont{Marchetti}},
  \bibinfo{journal}{Physical review letters} \textbf{\bibinfo{volume}{108}},
  \bibinfo{pages}{235702} (\bibinfo{year}{2012}).

\bibitem[{\citenamefont{Redner et~al.}(2013)\citenamefont{Redner, Hagan, and
  Baskaran}}]{redner2013structure}
\bibinfo{author}{\bibfnamefont{G.~S.} \bibnamefont{Redner}},
  \bibinfo{author}{\bibfnamefont{M.~F.} \bibnamefont{Hagan}}, \bibnamefont{and}
  \bibinfo{author}{\bibfnamefont{A.}~\bibnamefont{Baskaran}},
  \bibinfo{journal}{Biophysical Journal} \textbf{\bibinfo{volume}{104}},
  \bibinfo{pages}{640a} (\bibinfo{year}{2013}).

\bibitem[{\citenamefont{Tailleur and Cates}(2008)}]{tailleur2008statistical}
\bibinfo{author}{\bibfnamefont{J.}~\bibnamefont{Tailleur}} \bibnamefont{and}
  \bibinfo{author}{\bibfnamefont{M.~E.} \bibnamefont{Cates}},
  \bibinfo{journal}{Physical review letters} \textbf{\bibinfo{volume}{100}},
  \bibinfo{pages}{218103} (\bibinfo{year}{2008}).

\bibitem[{\citenamefont{Slowman et~al.}(2016)\citenamefont{Slowman, Evans, and
  Blythe}}]{slowman2016jamming}
\bibinfo{author}{\bibfnamefont{A.}~\bibnamefont{Slowman}},
  \bibinfo{author}{\bibfnamefont{M.}~\bibnamefont{Evans}}, \bibnamefont{and}
  \bibinfo{author}{\bibfnamefont{R.}~\bibnamefont{Blythe}},
  \bibinfo{journal}{Physical review letters} \textbf{\bibinfo{volume}{116}},
  \bibinfo{pages}{218101} (\bibinfo{year}{2016}).

\bibitem[{\citenamefont{Cates and Tailleur}(2013)}]{cates2013active}
\bibinfo{author}{\bibfnamefont{M.~E.} \bibnamefont{Cates}} \bibnamefont{and}
  \bibinfo{author}{\bibfnamefont{J.}~\bibnamefont{Tailleur}},
  \bibinfo{journal}{Europhysics Letters} \textbf{\bibinfo{volume}{101}},
  \bibinfo{pages}{20010} (\bibinfo{year}{2013}).

\bibitem[{\citenamefont{Sep{\'u}lveda and
  Soto}(2016)}]{sepulveda2016coarsening}
\bibinfo{author}{\bibfnamefont{N.}~\bibnamefont{Sep{\'u}lveda}}
  \bibnamefont{and} \bibinfo{author}{\bibfnamefont{R.}~\bibnamefont{Soto}},
  \bibinfo{journal}{Physical Review E} \textbf{\bibinfo{volume}{94}},
  \bibinfo{pages}{022603} (\bibinfo{year}{2016}).

\bibitem[{\citenamefont{Levis and Berthier}(2014)}]{levis2014clustering}
\bibinfo{author}{\bibfnamefont{D.}~\bibnamefont{Levis}} \bibnamefont{and}
  \bibinfo{author}{\bibfnamefont{L.}~\bibnamefont{Berthier}},
  \bibinfo{journal}{Physical Review E} \textbf{\bibinfo{volume}{89}},
  \bibinfo{pages}{062301} (\bibinfo{year}{2014}).

\bibitem[{\citenamefont{Stenhammar et~al.}(2014)\citenamefont{Stenhammar,
  Marenduzzo, Allen, and Cates}}]{stenhammar2014phase}
\bibinfo{author}{\bibfnamefont{J.}~\bibnamefont{Stenhammar}},
  \bibinfo{author}{\bibfnamefont{D.}~\bibnamefont{Marenduzzo}},
  \bibinfo{author}{\bibfnamefont{R.~J.} \bibnamefont{Allen}}, \bibnamefont{and}
  \bibinfo{author}{\bibfnamefont{M.~E.} \bibnamefont{Cates}},
  \bibinfo{journal}{Soft matter} \textbf{\bibinfo{volume}{10}},
  \bibinfo{pages}{1489} (\bibinfo{year}{2014}).

\bibitem[{\citenamefont{Mognetti et~al.}(2013)\citenamefont{Mognetti,
  {\v{S}}ari{\'c}, Angioletti-Uberti, Cacciuto, Valeriani, and
  Frenkel}}]{mognetti2013living}
\bibinfo{author}{\bibfnamefont{B.~M.} \bibnamefont{Mognetti}},
  \bibinfo{author}{\bibfnamefont{A.}~\bibnamefont{{\v{S}}ari{\'c}}},
  \bibinfo{author}{\bibfnamefont{S.}~\bibnamefont{Angioletti-Uberti}},
  \bibinfo{author}{\bibfnamefont{A.}~\bibnamefont{Cacciuto}},
  \bibinfo{author}{\bibfnamefont{C.}~\bibnamefont{Valeriani}},
  \bibnamefont{and} \bibinfo{author}{\bibfnamefont{D.}~\bibnamefont{Frenkel}},
  \bibinfo{journal}{Physical review letters} \textbf{\bibinfo{volume}{111}},
  \bibinfo{pages}{245702} (\bibinfo{year}{2013}).

\bibitem[{\citenamefont{Suma et~al.}(2014)\citenamefont{Suma, Gonnella,
  Marenduzzo, and Orlandini}}]{suma2014motility}
\bibinfo{author}{\bibfnamefont{A.}~\bibnamefont{Suma}},
  \bibinfo{author}{\bibfnamefont{G.}~\bibnamefont{Gonnella}},
  \bibinfo{author}{\bibfnamefont{D.}~\bibnamefont{Marenduzzo}},
  \bibnamefont{and}
  \bibinfo{author}{\bibfnamefont{E.}~\bibnamefont{Orlandini}},
  \bibinfo{journal}{Europhysics Letters} \textbf{\bibinfo{volume}{108}},
  \bibinfo{pages}{56004} (\bibinfo{year}{2014}).

\bibitem[{\citenamefont{Cates et~al.}(2010)\citenamefont{Cates, Marenduzzo,
  Pagonabarraga, and Tailleur}}]{cates2010arrested}
\bibinfo{author}{\bibfnamefont{M.~E.} \bibnamefont{Cates}},
  \bibinfo{author}{\bibfnamefont{D.}~\bibnamefont{Marenduzzo}},
  \bibinfo{author}{\bibfnamefont{I.}~\bibnamefont{Pagonabarraga}},
  \bibnamefont{and} \bibinfo{author}{\bibfnamefont{J.}~\bibnamefont{Tailleur}},
  \bibinfo{journal}{Proceedings of the National Academy of Sciences}
  \textbf{\bibinfo{volume}{107}}, \bibinfo{pages}{11715}
  (\bibinfo{year}{2010}).

\bibitem[{\citenamefont{Geyer et~al.}(2019)\citenamefont{Geyer, Martin,
  Tailleur, and Bartolo}}]{geyer2019freezing}
\bibinfo{author}{\bibfnamefont{D.}~\bibnamefont{Geyer}},
  \bibinfo{author}{\bibfnamefont{D.}~\bibnamefont{Martin}},
  \bibinfo{author}{\bibfnamefont{J.}~\bibnamefont{Tailleur}}, \bibnamefont{and}
  \bibinfo{author}{\bibfnamefont{D.}~\bibnamefont{Bartolo}},
  \bibinfo{journal}{Physical Review X} \textbf{\bibinfo{volume}{9}},
  \bibinfo{pages}{031043} (\bibinfo{year}{2019}).

\bibitem[{\citenamefont{Solon et~al.}(2015)\citenamefont{Solon, Caussin,
  Bartolo, Chat{\'e}, and Tailleur}}]{solon2015pattern}
\bibinfo{author}{\bibfnamefont{A.~P.} \bibnamefont{Solon}},
  \bibinfo{author}{\bibfnamefont{J.-B.} \bibnamefont{Caussin}},
  \bibinfo{author}{\bibfnamefont{D.}~\bibnamefont{Bartolo}},
  \bibinfo{author}{\bibfnamefont{H.}~\bibnamefont{Chat{\'e}}},
  \bibnamefont{and} \bibinfo{author}{\bibfnamefont{J.}~\bibnamefont{Tailleur}},
  \bibinfo{journal}{Physical Review E} \textbf{\bibinfo{volume}{92}},
  \bibinfo{pages}{062111} (\bibinfo{year}{2015}).

\bibitem[{\citenamefont{Marchetti et~al.}(2013)\citenamefont{Marchetti, Joanny,
  Ramaswamy, Liverpool, Prost, Rao, and Simha}}]{marchetti2013hydrodynamics}
\bibinfo{author}{\bibfnamefont{M.~C.} \bibnamefont{Marchetti}},
  \bibinfo{author}{\bibfnamefont{J.-F.} \bibnamefont{Joanny}},
  \bibinfo{author}{\bibfnamefont{S.}~\bibnamefont{Ramaswamy}},
  \bibinfo{author}{\bibfnamefont{T.~B.} \bibnamefont{Liverpool}},
  \bibinfo{author}{\bibfnamefont{J.}~\bibnamefont{Prost}},
  \bibinfo{author}{\bibfnamefont{M.}~\bibnamefont{Rao}}, \bibnamefont{and}
  \bibinfo{author}{\bibfnamefont{R.~A.} \bibnamefont{Simha}},
  \bibinfo{journal}{Reviews of modern physics} \textbf{\bibinfo{volume}{85}},
  \bibinfo{pages}{1143} (\bibinfo{year}{2013}).

\bibitem[{\citenamefont{Bialk{\'e} et~al.}(2015)\citenamefont{Bialk{\'e},
  Speck, and L{\"o}wen}}]{bialke2015active}
\bibinfo{author}{\bibfnamefont{J.}~\bibnamefont{Bialk{\'e}}},
  \bibinfo{author}{\bibfnamefont{T.}~\bibnamefont{Speck}}, \bibnamefont{and}
  \bibinfo{author}{\bibfnamefont{H.}~\bibnamefont{L{\"o}wen}},
  \bibinfo{journal}{Journal of Non-Crystalline Solids}
  \textbf{\bibinfo{volume}{407}}, \bibinfo{pages}{367} (\bibinfo{year}{2015}).

\bibitem[{\citenamefont{Soto and Golestanian}(2014)}]{soto2014run}
\bibinfo{author}{\bibfnamefont{R.}~\bibnamefont{Soto}} \bibnamefont{and}
  \bibinfo{author}{\bibfnamefont{R.}~\bibnamefont{Golestanian}},
  \bibinfo{journal}{Physical Review E} \textbf{\bibinfo{volume}{89}},
  \bibinfo{pages}{012706} (\bibinfo{year}{2014}).

\bibitem[{\citenamefont{Barberis and Peruani}(2019)}]{barberis2019phase}
\bibinfo{author}{\bibfnamefont{L.}~\bibnamefont{Barberis}} \bibnamefont{and}
  \bibinfo{author}{\bibfnamefont{F.}~\bibnamefont{Peruani}},
  \bibinfo{journal}{The Journal of chemical physics}
  \textbf{\bibinfo{volume}{150}} (\bibinfo{year}{2019}).

\bibitem[{\citenamefont{Guti{\'e}rrez et~al.}(2021)\citenamefont{Guti{\'e}rrez,
  Vanhille-Campos, Alarcon, Pagonabarraga, Brito, and
  Valeriani}}]{gutierrez2021collective}
\bibinfo{author}{\bibfnamefont{C.~M.~B.} \bibnamefont{Guti{\'e}rrez}},
  \bibinfo{author}{\bibfnamefont{C.}~\bibnamefont{Vanhille-Campos}},
  \bibinfo{author}{\bibfnamefont{F.}~\bibnamefont{Alarcon}},
  \bibinfo{author}{\bibfnamefont{I.}~\bibnamefont{Pagonabarraga}},
  \bibinfo{author}{\bibfnamefont{R.}~\bibnamefont{Brito}}, \bibnamefont{and}
  \bibinfo{author}{\bibfnamefont{C.}~\bibnamefont{Valeriani}},
  \bibinfo{journal}{Soft Matter} \textbf{\bibinfo{volume}{17}},
  \bibinfo{pages}{10479} (\bibinfo{year}{2021}).

\bibitem[{\citenamefont{Dandekar et~al.}(2020)\citenamefont{Dandekar,
  Chakraborti, and Rajesh}}]{dandekar2020hard}
\bibinfo{author}{\bibfnamefont{R.}~\bibnamefont{Dandekar}},
  \bibinfo{author}{\bibfnamefont{S.}~\bibnamefont{Chakraborti}},
  \bibnamefont{and} \bibinfo{author}{\bibfnamefont{R.}~\bibnamefont{Rajesh}},
  \bibinfo{journal}{Physical Review E} \textbf{\bibinfo{volume}{102}},
  \bibinfo{pages}{062111} (\bibinfo{year}{2020}).

\bibitem[{\citenamefont{Berg}(2004)}]{berg2004coli}
\bibinfo{author}{\bibfnamefont{H.~C.} \bibnamefont{Berg}},
  \emph{\bibinfo{title}{E. coli in Motion}} (\bibinfo{publisher}{Springer},
  \bibinfo{year}{2004}).

\bibitem[{\citenamefont{Dev and Chatterjee}(2018)}]{dev2018optimal}
\bibinfo{author}{\bibfnamefont{S.}~\bibnamefont{Dev}} \bibnamefont{and}
  \bibinfo{author}{\bibfnamefont{S.}~\bibnamefont{Chatterjee}},
  \bibinfo{journal}{Physical Review E} \textbf{\bibinfo{volume}{97}},
  \bibinfo{pages}{032420} (\bibinfo{year}{2018}).

\bibitem[{\citenamefont{Mandal and Chatterjee}(2021)}]{mandal2021effect}
\bibinfo{author}{\bibfnamefont{S.~D.} \bibnamefont{Mandal}} \bibnamefont{and}
  \bibinfo{author}{\bibfnamefont{S.}~\bibnamefont{Chatterjee}},
  \bibinfo{journal}{Physical Review E} \textbf{\bibinfo{volume}{103}},
  \bibinfo{pages}{L030401} (\bibinfo{year}{2021}).

\bibitem[{\citenamefont{Eisenbach et~al.}(2004)\citenamefont{Eisenbach, Tamada,
  Omann, Segall, Firtel, Meili, Gutnick, Varon, Lengeler, and
  Murakami}}]{eisenbach2004chemotaxis}
\bibinfo{author}{\bibfnamefont{M.}~\bibnamefont{Eisenbach}},
  \bibinfo{author}{\bibfnamefont{A.}~\bibnamefont{Tamada}},
  \bibinfo{author}{\bibfnamefont{G.}~\bibnamefont{Omann}},
  \bibinfo{author}{\bibfnamefont{J.}~\bibnamefont{Segall}},
  \bibinfo{author}{\bibfnamefont{R.}~\bibnamefont{Firtel}},
  \bibinfo{author}{\bibfnamefont{R.}~\bibnamefont{Meili}},
  \bibinfo{author}{\bibfnamefont{D.}~\bibnamefont{Gutnick}},
  \bibinfo{author}{\bibfnamefont{M.}~\bibnamefont{Varon}},
  \bibinfo{author}{\bibfnamefont{J.~W.} \bibnamefont{Lengeler}},
  \bibnamefont{and} \bibinfo{author}{\bibfnamefont{F.}~\bibnamefont{Murakami}},
  \emph{\bibinfo{title}{Chemotaxis}} (\bibinfo{publisher}{World Scientific
  Publishing Company}, \bibinfo{year}{2004}).

\bibitem[{\citenamefont{Chatterjee et~al.}(2014)\citenamefont{Chatterjee,
  Chatterjee, Pradhan, and Manna}}]{chatterjee2014interacting}
\bibinfo{author}{\bibfnamefont{R.}~\bibnamefont{Chatterjee}},
  \bibinfo{author}{\bibfnamefont{S.}~\bibnamefont{Chatterjee}},
  \bibinfo{author}{\bibfnamefont{P.}~\bibnamefont{Pradhan}}, \bibnamefont{and}
  \bibinfo{author}{\bibfnamefont{S.}~\bibnamefont{Manna}},
  \bibinfo{journal}{Physical Review E} \textbf{\bibinfo{volume}{89}},
  \bibinfo{pages}{022138} (\bibinfo{year}{2014}).

\bibitem[{\citenamefont{Chatterjee et~al.}(2016)\citenamefont{Chatterjee,
  Chatterjee, and Pradhan}}]{chatterjee2016symmetric}
\bibinfo{author}{\bibfnamefont{R.}~\bibnamefont{Chatterjee}},
  \bibinfo{author}{\bibfnamefont{S.}~\bibnamefont{Chatterjee}},
  \bibnamefont{and} \bibinfo{author}{\bibfnamefont{P.}~\bibnamefont{Pradhan}},
  \bibinfo{journal}{Physical Review E} \textbf{\bibinfo{volume}{93}},
  \bibinfo{pages}{062124} (\bibinfo{year}{2016}).

\bibitem[{\citenamefont{Das et~al.}(2023)\citenamefont{Das, Pradhan, and
  Chatterjee}}]{das2023optimum}
\bibinfo{author}{\bibfnamefont{D.}~\bibnamefont{Das}},
  \bibinfo{author}{\bibfnamefont{P.}~\bibnamefont{Pradhan}}, \bibnamefont{and}
  \bibinfo{author}{\bibfnamefont{S.}~\bibnamefont{Chatterjee}},
  \bibinfo{journal}{Physical Review E} \textbf{\bibinfo{volume}{108}},
  \bibinfo{pages}{034107} (\bibinfo{year}{2023}).

\bibitem[{\citenamefont{Nagar et~al.}(2008)\citenamefont{Nagar, Ha, and
  Park}}]{nagar2008boundary}
\bibinfo{author}{\bibfnamefont{A.}~\bibnamefont{Nagar}},
  \bibinfo{author}{\bibfnamefont{M.}~\bibnamefont{Ha}}, \bibnamefont{and}
  \bibinfo{author}{\bibfnamefont{H.}~\bibnamefont{Park}},
  \bibinfo{journal}{Physical Review E} \textbf{\bibinfo{volume}{77}},
  \bibinfo{pages}{061118} (\bibinfo{year}{2008}).

\bibitem[{\citenamefont{Chatterjee and
  Hayakawa}(2023)}]{chatterjee2023counterflow}
\bibinfo{author}{\bibfnamefont{A.~K.} \bibnamefont{Chatterjee}}
  \bibnamefont{and} \bibinfo{author}{\bibfnamefont{H.}~\bibnamefont{Hayakawa}},
  \bibinfo{journal}{Physical Review E} \textbf{\bibinfo{volume}{107}},
  \bibinfo{pages}{054905} (\bibinfo{year}{2023}).

\bibitem[{\citenamefont{Chacko et~al.}(2024)\citenamefont{Chacko, Muhuri, and
  Tripathy}}]{chacko2024clustering}
\bibinfo{author}{\bibfnamefont{J.}~\bibnamefont{Chacko}},
  \bibinfo{author}{\bibfnamefont{S.}~\bibnamefont{Muhuri}}, \bibnamefont{and}
  \bibinfo{author}{\bibfnamefont{G.}~\bibnamefont{Tripathy}},
  \bibinfo{journal}{Indian Journal of Physics} \textbf{\bibinfo{volume}{98}},
  \bibinfo{pages}{1553} (\bibinfo{year}{2024}).

\bibitem[{\citenamefont{Berg and Turner}(1990)}]{berg1990chemotaxis}
\bibinfo{author}{\bibfnamefont{H.~C.} \bibnamefont{Berg}} \bibnamefont{and}
  \bibinfo{author}{\bibfnamefont{L.}~\bibnamefont{Turner}},
  \bibinfo{journal}{Biophysical Journal} \textbf{\bibinfo{volume}{58}},
  \bibinfo{pages}{919} (\bibinfo{year}{1990}).

\bibitem[{\citenamefont{Binz et~al.}(2010)\citenamefont{Binz, Lee, Edwards, and
  Nicolau}}]{binz2010motility}
\bibinfo{author}{\bibfnamefont{M.}~\bibnamefont{Binz}},
  \bibinfo{author}{\bibfnamefont{A.~P.} \bibnamefont{Lee}},
  \bibinfo{author}{\bibfnamefont{C.}~\bibnamefont{Edwards}}, \bibnamefont{and}
  \bibinfo{author}{\bibfnamefont{D.~V.} \bibnamefont{Nicolau}},
  \bibinfo{journal}{Microelectronic Engineering} \textbf{\bibinfo{volume}{87}},
  \bibinfo{pages}{810} (\bibinfo{year}{2010}).

\bibitem[{\citenamefont{Li et~al.}(2017)\citenamefont{Li, Cai, Zhang, Si,
  Ouyang, Luo, and Tu}}]{li2017barrier}
\bibinfo{author}{\bibfnamefont{Z.}~\bibnamefont{Li}},
  \bibinfo{author}{\bibfnamefont{Q.}~\bibnamefont{Cai}},
  \bibinfo{author}{\bibfnamefont{X.}~\bibnamefont{Zhang}},
  \bibinfo{author}{\bibfnamefont{G.}~\bibnamefont{Si}},
  \bibinfo{author}{\bibfnamefont{Q.}~\bibnamefont{Ouyang}},
  \bibinfo{author}{\bibfnamefont{C.}~\bibnamefont{Luo}}, \bibnamefont{and}
  \bibinfo{author}{\bibfnamefont{Y.}~\bibnamefont{Tu}},
  \bibinfo{journal}{Physical review letters} \textbf{\bibinfo{volume}{118}},
  \bibinfo{pages}{098101} (\bibinfo{year}{2017}).

\bibitem[{\citenamefont{Silverman et~al.}(1976)\citenamefont{Silverman,
  Matsumura, and Simon}}]{silverman1976identification}
\bibinfo{author}{\bibfnamefont{M.}~\bibnamefont{Silverman}},
  \bibinfo{author}{\bibfnamefont{P.}~\bibnamefont{Matsumura}},
  \bibnamefont{and} \bibinfo{author}{\bibfnamefont{M.}~\bibnamefont{Simon}},
  \bibinfo{journal}{Proceedings of the National Academy of Sciences}
  \textbf{\bibinfo{volume}{73}}, \bibinfo{pages}{3126} (\bibinfo{year}{1976}).

\bibitem[{\citenamefont{Parkinson}(1976)}]{parkinson1976chea}
\bibinfo{author}{\bibfnamefont{J.~S.} \bibnamefont{Parkinson}},
  \bibinfo{journal}{Journal of bacteriology} \textbf{\bibinfo{volume}{126}},
  \bibinfo{pages}{758} (\bibinfo{year}{1976}).

\bibitem[{\citenamefont{Parkinson and Houts}(1982)}]{parkinson1982isolation}
\bibinfo{author}{\bibfnamefont{J.~S.} \bibnamefont{Parkinson}}
  \bibnamefont{and} \bibinfo{author}{\bibfnamefont{S.~E.} \bibnamefont{Houts}},
  \bibinfo{journal}{Journal of bacteriology} \textbf{\bibinfo{volume}{151}},
  \bibinfo{pages}{106} (\bibinfo{year}{1982}).

\bibitem[{\citenamefont{Parkinson and Parker}(1979)}]{parkinson1979interaction}
\bibinfo{author}{\bibfnamefont{J.~S.} \bibnamefont{Parkinson}}
  \bibnamefont{and} \bibinfo{author}{\bibfnamefont{S.~R.}
  \bibnamefont{Parker}}, \bibinfo{journal}{Proceedings of the National Academy
  of Sciences} \textbf{\bibinfo{volume}{76}}, \bibinfo{pages}{2390}
  (\bibinfo{year}{1979}).

\bibitem[{\citenamefont{Sanna and Simon}(1996)}]{sanna1996vivo}
\bibinfo{author}{\bibfnamefont{M.~G.} \bibnamefont{Sanna}} \bibnamefont{and}
  \bibinfo{author}{\bibfnamefont{M.~I.} \bibnamefont{Simon}},
  \bibinfo{journal}{Journal of bacteriology} \textbf{\bibinfo{volume}{178}},
  \bibinfo{pages}{6275} (\bibinfo{year}{1996}).

\end{thebibliography}
\end{document}